\documentclass[sigconf,screen]{acmart}

\AtBeginDocument{%
  }

\copyrightyear{2026}
\acmYear{2026}
\setcopyright{cc}
\setcctype{by}
\acmConference[CHI '26]{Proceedings of the 2026 CHI Conference on Human Factors in Computing Systems}{April 13--17, 2026}{Barcelona, Spain}
\acmBooktitle{Proceedings of the 2026 CHI Conference on Human Factors in Computing Systems (CHI '26), April 13--17, 2026, Barcelona, Spain}
\acmPrice{}
\acmDOI{10.1145/3772318.3790505}
\acmISBN{979-8-4007-2278-3/2026/04}

\graphicspath{{figures/}{pictures/}{images/}{./}} 

\usepackage{makecell}
\usepackage{amsmath}
\usepackage{algorithm}

\usepackage{color}
\usepackage{bbding}
\usepackage{booktabs}
\usepackage{hhline}

\definecolor{shellypurple}{HTML}{930FC1}

\usepackage{enumitem}
\setlist{noitemsep,parsep=0pt,partopsep=0pt, leftmargin=10pt} 
\usepackage{listings}
\usepackage{multirow}
\usepackage{colortbl}

\AtBeginDocument{

}
\newcommand{\eg}{{\it e.g.,\ }}
\newcommand{\etal}{{\it et al.\ }}
\newcommand{\etc}{{\it etc.}}
\newcommand{\ie}{{\it i.e.,\ }}

\newcommand{\nistart}[1]{{\noindent\textbf{#1.}}}

\definecolor{Restrict}{HTML}{eee0da}
\definecolor{Expand}{HTML}{d3e5ef}
\definecolor{Refine}{HTML}{dbeddb}
\definecolor{Organize}{HTML}{e8deee}
\definecolor{Follow-up}{HTML}{fdecc8}
\definecolor{Initial}{HTML}{ffe2dd}
\definecolor{Human}{HTML}{fadec9}
\definecolor{AI}{HTML}{d3e5ef}
\definecolor{Direct}{HTML}{eee0da}
\definecolor{Indirect}{HTML}{dbeddb}
\definecolor{Visual Brush Selection}{HTML}{f5e0e9}
\definecolor{Widget Selection}{HTML}{fdecc8}
\definecolor{Inline Annotation}{HTML}{dbeddb}
\definecolor{Interactive Visual Click}{HTML}{ffe2dd}
\definecolor{Flowchart Manipulation}{HTML}{e8deee}
\definecolor{Tree Manipulation}{HTML}{fadec9}
\definecolor{Text Brush Selection}{HTML}{d3e5ef}
\definecolor{Text Input or Editing}{HTML}{ffffcc}
\definecolor{Inline Highlighting}{HTML}{e9fbe0}
\definecolor{Viewpoint Navigation}{HTML}{e8deee}
\definecolor{Sketch}{HTML}{c8e3f3}
\definecolor{Spreadsheet Manipulation}{HTML}{f1f0ef}
\definecolor{Speech}{HTML}{f1f0ef}
\definecolor{Global Software Manipulation}{HTML}{f1f0ef}
\definecolor{Drag and Drop}{HTML}{f1f0ef}

\definecolor{tablerowcolor}{rgb}{0.667,0.667,0.667 }
\definecolor{tablerowcolor2}{rgb}{0,0,0}
\definecolor{visual}{HTML}{e8efd9}
\definecolor{motion}{HTML}{fde7d5}
\definecolor{narrative}{HTML}{e2dce9}
\definecolor{audio}{HTML}{d6ebf2}
\definecolor{bluecrayola}{rgb}{0.12,0.46,1.0}
\definecolor{arti}{HTML}{ffb900}

\newcommand{\revise}[1]{{\color{black} #1}}
\newcommand{\review}[1]{\textcolor{purple}{}}
\definecolor{copied}{RGB}{165, 165, 165} 

\definecolor{wyf}{RGB}{237, 85, 106} 

\newcommand{\re}[1]{\textcolor{black}{#1}}

\begin{document}
\title[Modeling the Synergy of Prompts and Interactions in Human-GenAI Collaboration]{Interaction-Augmented Instruction: Modeling the Synergy of Prompts and Interactions in Human-GenAI Collaboration}

\author{Leixian Shen}
\authornote{Work done during internship at Microsoft Research Asia.}
\orcid{0000-0003-1084-4912}
\affiliation{%
  \institution{The Hong Kong University of Science and Technology}
  \city{Hong Kong SAR}
  \country{China}
}
\email{lshenaj@connect.ust.hk}

\author{Yifang Wang}
\orcid{0000-0001-6267-9440}
\affiliation{%
  \institution{Florida State University}
  \city{Tallahassee}
  \state{Florida}
  \country{United States}
}
\email{yifang.wang@fsu.edu}

\author{Huamin Qu}
\orcid{0000-0002-3344-9694}
\affiliation{%
  \institution{The Hong Kong University of Science and Technology}
  \city{Hong Kong SAR}
  \country{China}
}
\email{huamin@cse.ust.hk}

\author{Xing Xie}
\orcid{0000-0002-8608-8482}
\affiliation{%
  \institution{Microsoft Research Asia}
  \city{Beijing}
  \country{China}
}
\email{xingx@microsoft.com}

\author{Haotian Li}
\authornote{Haotian Li is the corresponding author.}
\orcid{0000-0001-9547-3449}
\affiliation{
  \institution{Microsoft Research Asia}
  \city{Beijing}
  \country{China}
}
\email{haotian.li@microsoft.com}

\begin{abstract}
Text prompt is the most common way for human-generative AI (GenAI) communication.
Though convenient, it is challenging to convey fine-grained and referential intent.
One promising solution is to combine text prompts with precise GUI interactions, like brushing and clicking.
However, there lacks a formal model to capture synergistic designs between prompts and interactions, hindering their comparison and innovation.
To fill this gap, via an iterative and deductive process, we develop the Interaction-Augmented Instruction (IAI) model, a compact entity–relation graph 
formalizing how the combination of interactions and text prompts enhances human-GenAI communication.
With the model, we distill twelve recurring and composable atomic interaction paradigms from prior tools, verifying our model's capability to facilitate systematic design characterization and comparison.
Four usage scenarios further demonstrate the model's utility in applying, refining, and innovating these paradigms. 
These results illustrate the IAI model's descriptive, discriminative, and generative power for shaping future GenAI systems.
\end{abstract}



\begin{teaserfigure}
\centering
\includegraphics[width=\linewidth]{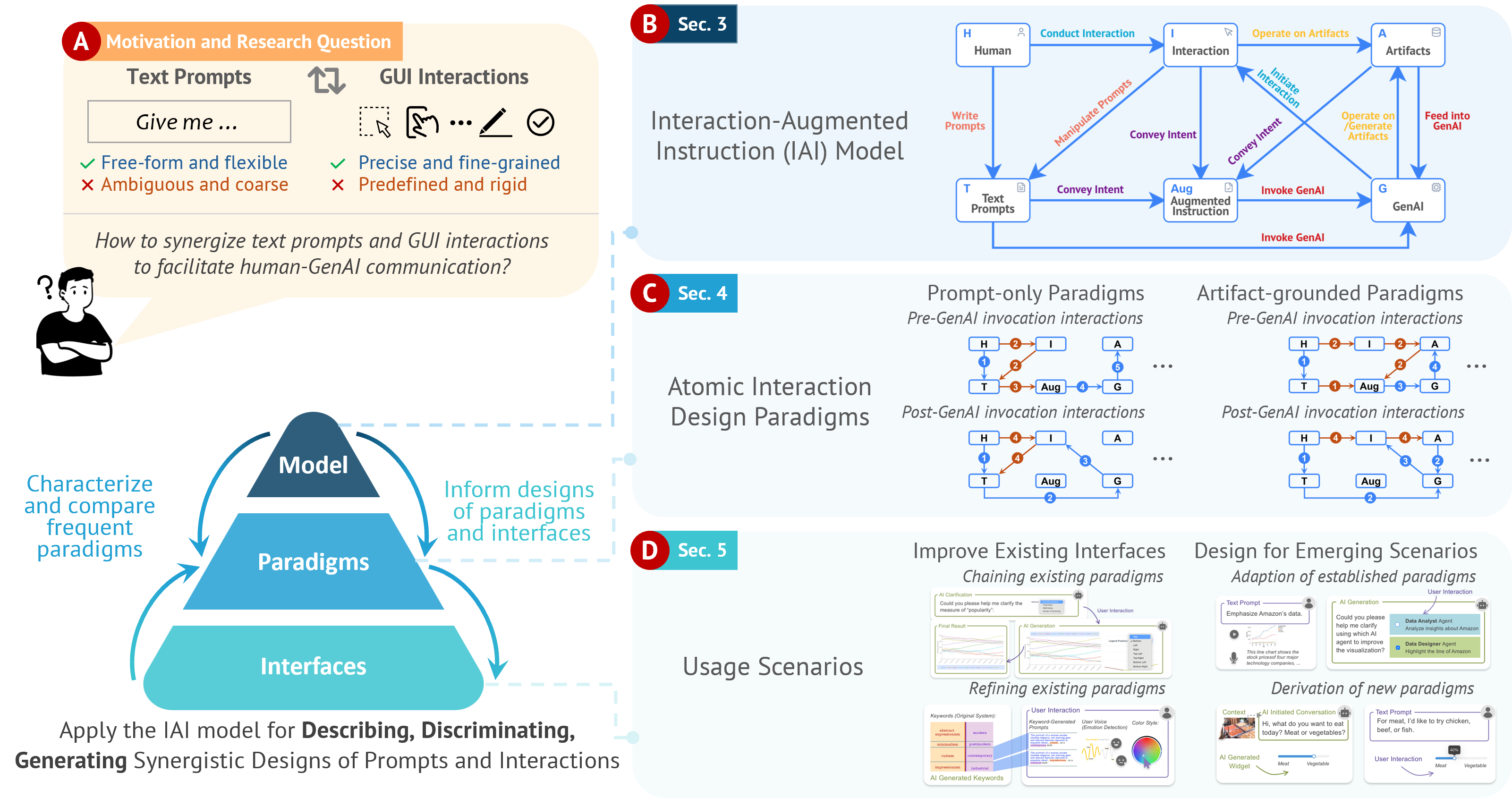}
\vspace{-15px}
\caption{
This paper addresses the research question of \textbf{(A)} how to synergize text prompts and GUI interactions to facilitate human–GenAI communication. We propose the Interaction-Augmented Instruction (IAI) model \textbf{(B)}. The model enables systematic characterization and comparison of existing paradigms \textbf{(C)} and guides the design of new interfaces \textbf{(D)}, demonstrating its \textit{descriptive, discriminative, and generative power} for shaping future GenAI systems. 
}
\label{fig: teaser}
\end{teaserfigure}

\maketitle

\section{Introduction}


Generative AI (GenAI) has rapidly become a general-purpose technology, enabling new intelligent applications across domains such as design~\cite{DVSurvey}, education~\cite{law2024application}, and data analysis~\cite{Priya}. Modern large language and multimodal models can interpret open-ended text prompts and generate diverse outputs (\eg text, image, video), making the text prompt interface the dominant paradigm for human–AI communication~\cite{NLISurvey}. 
Yet free-form text is often too ambiguous and coarse to express fine-grained or nuanced intent~\cite{Instructions, Riche2025, Subramonyam2023}.
For instance, in image editing, a user might want to modify ``\textit{the upper-right flower petal}'', but a text prompt alone is hard to accurately localize that target visual element. Likewise, in coding, asking ``\textit{write a loop to process this data}'' could imply different iteration strategies.
Correspondingly, in practice, users often struggle to articulate sufficient detail or to break down complex goals into a single prompt.


To address these limitations, a promising strategy is to combine general-purpose but imprecise text prompts with dedicated interactions (\eg clicks, drags, brushes, \etc) through graphical user interfaces (GUIs), as shown in \autoref{fig: teaser}-A. 
The synergy enables both flexible and fine-grained user control.
\review{Q1}\revise{
For example, consider Jason, a creative director designing a stylized magazine cover portrait. His intent is clear: ``\textit{I need a stylized portrait for a magazine cover, with high fashion and artistic lighting}''.
Compared with the basic prompt-only workflow (\autoref{fig:motivating-case}-A), this task can be realized through a spectrum of interface designs (\autoref{fig:motivating-case} B–F) that combine prompts and interactions in distinct ways.
Concretely, some interfaces let him refine or expand the text prompt before generation (\autoref{fig:motivating-case}-B), for instance, selecting ``\textit{artistic lighting}'' for fine-tuning or switching to a ``\textit{minimal black-and-white}'' style via UI controls. This functionality speeds up exploration but still assumes all intent can be expressed in language~\cite{Brade2023,Almeda2024}. 
To go beyond language, Jason might manually sketch a portrait outline or brush key regions, merging these non-linguistic signals with the prompt for more precise spatial and referential control (\autoref{fig:motivating-case}-C)~\cite{Lin2025a, Kim2023d}. 
When an initial render already exists, he may directly draw a necklace on the portrait, with the system incorporating both the prompt and the manipulated artifact content to enable precise, auditable edits grounded in the actual image (\autoref{fig:motivating-case}-D)~\cite{Masson2023b,Liu2024d}.
After generating an initial image, some systems propose interactive controls (\autoref{fig:motivating-case}-E), such as sliders for keyword (\eg``\textit{stylized}'' and ``\textit{high fashion}'') attention or color palettes for finer ``\textit{artistic lighting}'', allowing Jason to steer subsequent generations without rewriting text~\cite{Wang2024g,Gmeiner2025}. 
Others analyze the generated artifact and propose structured options (\autoref{fig:motivating-case}-F): Jason can choose to generate cover text, select the type, style, and length, and assemble a structured instruction for a magazine tagline~\cite{Zhang2023b,Kim2023c}. 

\begin{figure*}[t]
  \centering
    \includegraphics[width=\linewidth]{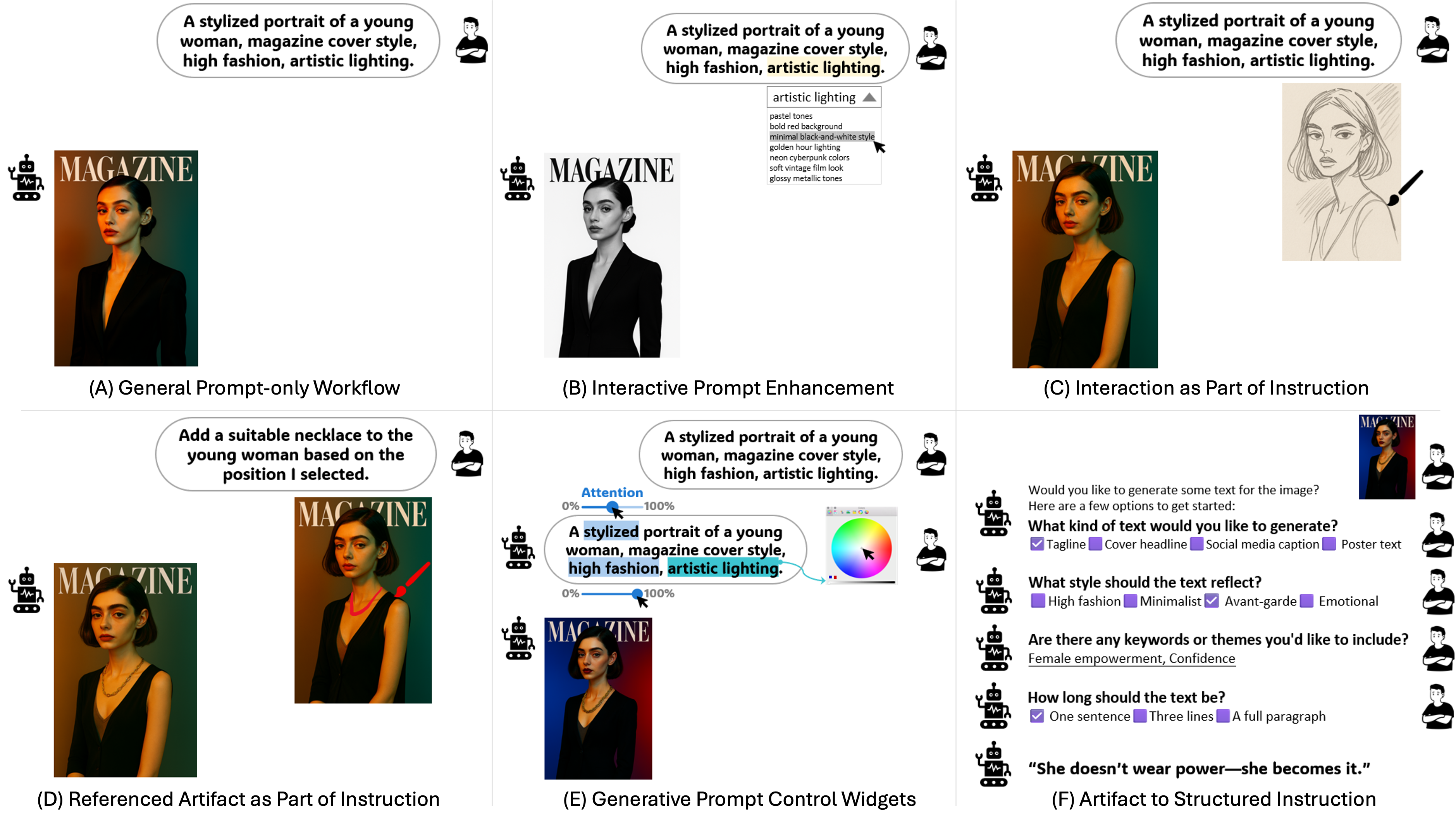}
     \vspace{-10px}
  \caption{\review{Q1}\revise{A unified image-generation task realized with distinct interaction-augmented designs: (A) prompt-only baseline; (B) prompt refinement via interaction; (C) combining prompt with non-linguistic interaction signals; (D) grounding instructions in artifact manipulation; (E) post-generation prompt control widgets; (F) artifact-anchored structured follow-ups.}}
  \label{fig:motivating-case}
\end{figure*}

Although these designs support the same creative goal, they differ fundamentally in various dimensions, such as where user intent is specified (\eg text, interaction, artifact), when interactions occur (pre- or post-model invocation), and what information is ultimately delivered to GenAI. 
Those structural differences affect referential fidelity, controllability, provenance, and user experience.
Furthermore, such strategies have not yet been formalized as a comprehensive \textit{interaction model} to guide related research and practices.}
As a pioneering study in the HCI field notes~\cite{Beaudouin-Lafon2004}, these ``point-like'' research demonstrations of the strategy are often insufficient due to the challenges of transforming task-specific demonstrations to broader real-world applications and understanding the advancement of related techniques.
It is essential to propose general interaction models to facilitate the careful examination of existing designs (\textit{descriptive power}), the comparison between design variations (\textit{discriminative power}), and the consequent innovation of new interaction designs for emerging tasks (\textit{generative power})~\cite{Beaudouin-Lafon2004, Beaudouin-Lafon2000, Liao2023}.

Towards this goal, there has been a series of previous research that attempted to reveal and advance the interaction-text prompt synergy.
Gao~\etal~\cite{Gao2024} and Hu~\etal~\cite{Cook2025} have developed taxonomies of interaction paradigms for interactive GenAI applications, such as GenAI as a medium, tool, partner, and mediator.
However, the taxonomy is mostly descriptive and is difficult to use to guide the generation of new designs for new scenarios.
Riche \etal~\cite{Riche2025} proposed AI-Instruments, extending instrumental interaction~\cite{Beaudouin-Lafon2000} by treating prompts themselves as manipulable interface objects, guided by three principles: Reification, Reflection, and Grounding. 
These principles enrich the considerations for generating new interaction designs but fall short in comparing multiple designs within a model that explicitly represents the interplay between text prompts and other interaction modalities.

To fill this gap, through a deductive and iterative process, we propose the Interaction-Augmented Instruction (IAI) model: a compact entity–relation graph that makes explicit how precise GUI interactions and free-form prompts are composed into the executable instructions consumed by GenAI (\autoref{fig: teaser}-B). 
The model comprises six entities: Human (\textbf{H}), Interaction (\textbf{I}), Text Prompt (\textbf{T}), Augmented Instruction (\textbf{Aug}), GenAI (\textbf{G}), and Artifact (\textbf{A}). 
By consolidating candidate roles into this minimal set, the model highlights \textbf{Aug} as the explicit input to GenAI, thereby simplifying comparison across paradigms and clarifying how different interfaces combine \textbf{T}, \textbf{I}, and \textbf{A} to construct \textbf{Aug}.
We further systematically enumerated all pairwise relations and preserved only those that are semantically meaningful, discriminative, and interpretable (\autoref{tab: entity}).



Building on this foundation, to examine if our IAI model can be generally applied to describe and differentiate existing GenAI systems' interfaces, we revisit a curated corpus of GenAI-powered interactive systems and map each system’s interaction workflow to one or more directed paradigm graphs (\autoref{fig: teaser}-C).  
From the corpus, we distill twelve frequent and composable atomic paradigms organized by interaction timing (pre- or post-GenAI invocation) and resource availability (prompt-only vs. artifact-grounded)\footnote{We provide an interactive browser for paradigms and annotated tools: \url{https://interaction-augmented-instruction.github.io/}}. 
These paradigm graphs based on the IAI model serve as high-level abstractions of interaction workflows, enabling tools to be clustered and compared.

The IAI model provides not only a theoretical lens but also a practical framework for future interface design. 
The twelve design paradigms derived from the IAI model can be applied, refined, and extended for creating new interfaces that meet diverse scenario demands.
Unexplored interaction design paradigms are also revealed from comparing the existing paradigms with the entire IAI model.
To demonstrate the power of generating new interface design and novel interaction paradigms, we present four usage scenarios that bridge conceptual models with actionable design processes (\autoref{fig: teaser}-D).
These usage scenarios show how the IAI model can guide the iterative enhancement of existing tools by chaining or editing existing design paradigms, while also supporting the adaptation of established paradigms and the derivation of new atomic paradigms for emerging scenarios.
Looking forward, we envision the IAI model as a foundation for advancing human–GenAI communication, opening new opportunities to design more transparent, controllable, and creative GenAI systems.

To summarize, our work makes three main contributions: 
\begin{itemize}
\item 
Interaction-Augmented Instruction model (\autoref{sec: model}): We propose a new interaction model formalizing how task-specific interactions and free-form prompts jointly generate instructions for GenAI, enabling richer human–AI collaboration.
\item 
Design paradigms (\autoref{sec: design paradigm}): We apply this model to existing interfaces by encoding each as directed paradigm graphs, deriving 12 recurring and composable atomic paradigms, enabling systematic design characterization and comparison.
\item 
Usage scenarios (\autoref{sec: case}): 
We use four usage scenarios to demonstrate how our model can improve existing interface design and generate new design paradigms for emerging scenarios. 
\end{itemize}

\section{Related Work}
Prior work on human–GenAI collaboration covers communication paradigms, design patterns, and theoretical models; we review these to highlight progress and motivate a unified model of interaction-augmented instructions.

\subsection{Human–GenAI Communication}
\label{sec:communication}

Human–GenAI communication spans a spectrum of interface strategies, from free-form text prompting to fully encapsulated tools and hybrid paradigms that combine prompting with GUI interactions. 


Free-form text prompts remain the dominant way for human–GenAI communication due to their flexible, expressive, and low-friction~\cite{Wu2022e,Chung2023}. However, prompt-only interaction has well-documented limitations: natural language is often ambiguous, underspecified, or ill-structured for task-specific operations, which increases iteration cost and reduces control~\cite{Riche2025,Subramonyam2023,Instructions}. Empirical studies highlight recurring challenges such as prompt formulation, disambiguation, and intent steering—for example, users struggle to localize visual elements precisely through language alone~\cite{Masson2023b,Fok2024} or to specify the exact semantic behavior desired in code edits~\cite{Feng2024,Zhu}. 

Another line of interface design has been to encapsulate GenAI capabilities within backend agents or domain-specific tools~\cite{Bansal}. In such designs, the system exposes a constrained interface (\eg widgets, templates, or menus)~\cite{lin2023inksight, Masson2025, Wang2025, WonderFlow} or totally automate the process while GenAI executes domain logic behind the scenes~\cite{dataplayer, datadirector, NarrativePlayer}. 
For example, Textoshop~\cite{Masson2025} allows text editing entirely through drawing software-like interactions, with prompts handled internally; fully automated agents similarly hide prompting from the user~\cite{datadirector,Zhao2024a}.
This lowers user burden but constrains openness and the expressive flexibility of prompting.

A more flexible and increasingly prevalent approach is to combine prompting with GUI interactions—what we call interaction-augmented instructions. Here, prompts are enriched by targeted user actions such as clicking, brushing, sketching, or selecting~\cite{Gao2024, Instructions, Cook2025, vistalk, Luo2025}. These interactions may occur before the GenAI invocation, \eg selecting elements to constrain the scope of a prompt~\cite{Xie2024a,Masson2023b,Wang2023a,Chen2024d,Fok2024,Liu2024d,Liu2023c,NotePlayer,Zhang2025a}, organizing multiple prompts into structured forms such as trees or graphs~\cite{Feng2024,Zhu, Wu2022e, Gmeiner2025}, sketching to guide content generation~\cite{Lin2025a,Kim2023d,Nguyen2023,Zhou2023}, annotating text to extend context or highlight specific information~\cite{dataplaywright}, or after it, \eg clarifying user intent through follow-up queries~\cite{Bursztyn2021,Baek2023,Andrews2024,Ma2024,Gmeiner2025}, presenting suggestions for human confirmation~\cite{Laban2023,Wu2023c,Angert2023, Gao2025}, converting prompt into structured GUI components~\cite{Zhang2025, Cai2024}, or enabling direct post-generation artifact manipulation~\cite{xie2025datawink,Vaithilingam2024,Suh2023,Gero2022,Petridis2024,Jiang2023}. Together, these paradigms illustrate diverse strategies for balancing the openness of prompting with the precision of interaction.

While many studies demonstrate the effectiveness of such hybrid paradigms in task-specific contexts, prior work often remains fragmented that do not generalize easily to broader real-world applications~\cite{Beaudouin-Lafon2004}. 
To advance the field, it is essential to propose general interaction models that can systematically capture these designs, enable structured comparison across paradigms, and inspire new interaction designs. Our work addresses this need by introducing a unifying formalism for describing, discriminating, and generating interaction-augmented instruction paradigms.

\subsection{Human–GenAI Interaction Models and Design Patterns}
\label{sec: interaction model}
Interaction design in modern UIs has evolved from early WIMP and direct-manipulation paradigms to richer post-WIMP models that foreground instruments and objects over low-level commands~\cite{VanDam1997,Lee2021c}.
Direct-manipulation interfaces emphasized immediate, visible control—users act on representations of domain objects and observe one-to-one effects~\cite{Shneiderman1983}. Building on this view, Beaudouin-Lafon’s Instrumental Interaction framed instruments as mediators between user actions and artifacts, where each instrumented action maps directly to a target operation~\cite{Beaudouin-Lafon2000, Beaudouin-Lafon2000a}. Subsequent work introduced ``substrates'' as ``places for interaction'', a mechanism to combine power and simplicity by structuring where and how instruments apply in complex interfaces~\cite{Mackay2025}.
These classical models are well suited to deterministic manipulation, but they assume a one-to-one relation between an interaction and its effect on an artifact. Generative systems violate that assumption: a single high-level instruction (\eg a free-form text prompt) can trigger multiple, nonlocal transformations across an artifact or even across multiple artifacts. This multiplicity and indirection require rethinking how instruments, intentions, and objects are modeled in an era of GenAI.

\review{Q2}\revise{
Recent HCI work has begun to explore this space. 
A series of surveys and taxonomies catalog interaction modalities and recurring patterns. For example, Lehmann~\etal~\cite{Lehmann2024} analyze how UI affordances mediate access to model capabilities, Luera~\etal~\cite{Luera} review input modalities across GenAI applications, Gao~\etal~\cite{Gao2024} propose a taxonomy of human–LLM communication modes, and Hu~\etal~\cite{Cook2025} identify general paradigms, \ie GenAI as medium, tool, partner, and mediator. 
While valuable as high-level accounts, these taxonomies provide only coarse-grained descriptive and discriminative power and offer limited generative power. For example, Gao~\etal~\cite{Gao2024} label a \textit{prompt decomposition} subcategory, which is named as ``UI for reasoning'' in their paper, but cannot distinguish whether decomposition is performed by the human~\cite{Wu2022e} or GenAI~\cite{Jiang2023}. 
Following Hu~\etal~\cite{Cook2025}, these systems are similarly grouped under ``GenAI as medium and UI affordance for exploration'', yet cannot differentiate whether interactions are initiated by the human~\cite{Brade2023} or GenAI~\cite{Suh2023}, and whether the human~\cite{Suh2023} or GenAI~\cite{Brade2023} ultimately operates on the artifact. 
Their main drawback is the lack of fine-grained characterization of interaction flow between humans and AI, leading to the insufficient descriptive and discriminative powers.
Shen~\etal~\cite{Instructions} advanced them by introducing detailed considerations of action timing and initiators beyond core purposes for interaction, \ie restricting, expanding, organizing, and refining prompts. 
However, it still falls short in distinguishing how intent is formulated (\eg text- vs. artifact-grounded) and what forms the post-interaction information takes (\eg linguistic vs. non-linguistic).
For example, Shen~\etal~\cite{Instructions} cannot characterize the difference between \autoref{fig:motivating-case}-E and \autoref{fig:motivating-case}-F as ``AI-initiated, post-invocation instruction extension'', though they provide different ways to facilitate user intent formulation with pure text prompts or existing artifacts.

Complementing these descriptive efforts, Riche~\etal~\cite{Riche2025} introduce AI-Instruments, extending instrumental interaction~\cite{Beaudouin-Lafon2000} to generative settings by making prompts themselves manipulable objects and articulating principles of \textit{reification}, \textit{reflection}, and \textit{grounding}. 
Though these high-level guidelines help consider how human-AI interaction should be designed,
they also suffer from limited and coarse-grained 
descriptive or discriminative power across diverse interfaces.
The main reason is their incapability in capturing system-internal information-flow differences, such as interaction timing, the explicit role of artifacts, or how composite instructions combine multiple modalities, as shown in \autoref{fig:motivating-case}.
\re{For example, Riche~\etal~\cite{Riche2025}  designed four interface prototypes according to the principles. 
In their work, ``Fillable Brushes'' leverage user interactions on existing artifacts and new prompts for next-step AI generation (similar to \autoref{fig:motivating-case}-D and P4 in our summarized paradigms in~\autoref{tab: paradigm}).
In ``Fragments'', AI can break down prompts and generate alternative options based on different design dimensions for users to interact with (similar to P6 in our summarized paradigms in~\autoref{tab: paradigm}).
They both satisfy the three principles, yet clearly correspond to different interaction flow designs that the model cannot structurally distinguish, such as the timing of interactions.}



The motivating example (\autoref{fig:motivating-case}) and the discussion above highlights the need for a unified, systematic model, which can precisely describe how prompts, interactions, and artifacts are combined across various dimensions like timing and resource axes; discriminate between closely related designs; and generate actionable blueprints for evolving real-world interfaces. The next section introduces such a model, addressing these gaps and enabling the descriptive, discriminative, and generative capabilities that prior frameworks lack.}

\section{Interaction-Augmented Instruction Model}
\label{sec: model}
To systematically capture the synergy between text prompts and interactions in GenAI system interfaces, we propose the \textit{Interaction-Augmented Instruction} Model, which is a formal, entity–relation graph that unifies general GUI interaction concepts with new constructs introduced by GenAI (\autoref{fig: graph}). 
It distills the essential components of human–GenAI communication and their relations, enabling descriptive, discriminative, and generative analysis of diverse interfaces.
This section first describes how we derive the key model concepts, \ie entities (\autoref{sec:model_entity}) and relations (\autoref{sec:model_relation}), and then discusses applying these concepts to represent an interface design with directed graphs (\autoref{sec:model_graph}).

\subsection{Entities}
\label{sec:model_entity}
The IAI Model comprises six core entities: \textbf{Human (H)}, \textbf{Interaction (I)}, \textbf{Artifact (A)}, \textbf{Text Prompt (T)}, \textbf{Augmented Instruction (Aug)}, and \textbf{Generative AI (G)}. Concretely, we treat an entity as a semantically coherent object or agent in the interaction ecology.
\autoref{tab: entity} provides an overview for all entities and relations.

We derived the entities through an iterative, deductive process that begins with asking a simple question: what are the irreducible elements that appear in the common prompt-driven and interaction-driven paths in practice?
Two empirically ubiquitous paths served as the starting point.
The first is the canonical \textit{prompt-driven flow}, H~$\rightarrow$~T~$\rightarrow$~G~$\rightarrow$~A, in which a \textbf{human (H)} composes a \textbf{text prompt (T)} that the \textbf{GenAI (G)} model executes to produce or modify an \textbf{artifact (A)}~\cite{Riche2025, Lee2025}. 
The second is the general \textit{GUI interaction flow},
H~$\rightarrow$~I~$\rightarrow$~A, in which a \textbf{human (H)} leverages \textbf{interactions (I)} on GUIs to act on an \textbf{artifact (A)}~\cite{Beaudouin-Lafon2000,Shneiderman1983}.
Based on the two interaction paths, three distinctions are critical to derive our model.

First, to capture the qualitative differences between modalities, we separate \textbf{Text Prompt (T)} from \textbf{Interaction (I)}. 
\review{Q3, Q5.1}\revise{Conceptually, \textbf{T} is a free-form, general-purpose, natural language specification of intent (\eg ``\textit{a stylized portrait of a young woman}'' in \autoref{fig:motivating-case}-A). 
\textbf{I} denotes focused, non-textual, often GUI user actions (\eg click in \autoref{fig:motivating-case}-B, sketch in \autoref{fig:motivating-case}-C, brush in \autoref{fig:motivating-case}-D, drag in \autoref{fig:motivating-case}-E, widget selection in \autoref{fig:motivating-case}-F) that can modify, supplement, or operationalize concrete referents or constraints (\eg a brush mask, a bounding box, a selected code block)}.
This separation matters because \textbf{T} and \textbf{I} differ sharply in expressivity, precision, and in how the model understands them.
However, such separation is not sufficient to represent what the model actually consumes in many practical GenAI workflows.


To address the issue, secondly, we introduce \textbf{Augmented Instruction (Aug)} as a new entity. In GenAI system interfaces, text prompts and interactions are not independent ``inputs'' that the model somehow interprets in isolation.
Interactions often produce structured, non-linguistic constraints (\eg pixel masks, coordinate ranges, AST node identifiers, selected table rows or text segments) that must be encoded, normalized, and attached to a prompt in a machine-readable form~\cite{dataplaywright, Chung2023, Liu2024d, Liu2023e, Almeda2024}. 
\review{Q3, Q4}\revise{Viewing \textbf{Aug} as the explicit input to GenAI simplifies paradigm comparisons, as tools differ mainly in how they combine \textbf{T}, \textbf{I}, and \textbf{A} to build \textbf{Aug}.}

Third, during iteration we consolidated several auxiliary entities (found in \autoref{sec: interaction model}) into the six core entities to keep the model parsimonious while maintaining expressiveness.
Concretely:
(1) \textbf{Context inputs} (\eg retrieved passages, prior conversation state, uploaded reference files) are modeled as part of Artifact (A) because they function as domain objects that ground instructions~\cite{Gao2024}. 
Their inclusion is modeled via A~$\rightarrow$~Aug (artifact-derived constraints) or A~$\rightarrow$~G (artifact supplied as raw model context).
(2) \textbf{Temporary interaction products} (\eg highlighted spans, sketched regions, intermediate masks, or annotation buffers) are treated as manifestations of Interaction (I) rather than independent entities~\cite{Cook2025}. They have only transient semantic life: they exist to constrain or point to artifacts or prompts and then are encoded into instructions or discarded. 
(3) \textbf{Widgets} have two aspects: when a widget is a UI primitive invoked by the human it is part of Interaction (H~$\rightarrow$~I)~\cite{Feng2024, Masson2023b}, and when it is generated by the model it is represented as an AI~$\rightarrow$~interaction proposal (G~$\rightarrow$~I) that the human may accept (I~$\rightarrow$~T~\cite{Andrews2024}, I~$\rightarrow$~Aug~\cite{Wang2024g}, or I~$\rightarrow$~A~\cite{Vaithilingam2024}). This treatment preserves provenance without inflating the entity set.
(4) \textbf{External tool} invocation is treated as part of GenAI’s internal behavior (G~$\rightarrow$~A). Many systems route domain-specific executors (\eg image editors, compilers, specialized APIs) behind the model; these are execution mechanisms rather than affecting interaction paradigm design~\cite{Instructions}. Modeling them as separate entities would conflate execution architecture with the interaction paradigms we aim to capture. 

Putting these together yields six entities (\autoref{tab: entity}), and each plays a distinct semantic role.
Conceptually the six-entity set is \emph{necessary}: removing any entity collapses an entire class of workflows, \eg without \textbf{I}, interaction-augmented cases degenerate to prompt-only; 
without \textbf{T}, prompt-only systems vanish;
without \textbf{Aug}, interaction-infused instructions cannot be distinguished from raw prompts; 
without \textbf{A}, there is no target object;  
without \textbf{G} or \textbf{H}, agency is undefined. 
They are also \emph{sufficient} to express human–GenAI communication paradigms with the interplay of interactions and prompts, as they are a universe of all nodes in both prompt-driven interaction flow and GUI interaction flow.
This point is further justified through revisiting existing research tools in \autoref{sec: design paradigm}.

\begin{table*}[t]
\centering
\renewcommand{\arraystretch}{1.3}

\caption{\re{Entities and Relations in the Interaction-Augmented Instruction Model. }
}

\definecolor{blue}{RGB}{33,113,181}
\definecolor{orange}{RGB}{230,85,13}
\definecolor{teal}{RGB}{0,128,128}
\definecolor{purple}{RGB}{118,42,131}
\definecolor{artifact}{RGB}{0,139,139}
\definecolor{red}{RGB}{203,24,29}

\label{tab: entity}

\begin{tabular}{|p{1.8cm}|p{6cm}|p{8.8cm}|}
\hline
\textbf{Entity} & \textbf{Description and Constraints} & \textbf{Linked Entities (Relation)} \\
\hline

\multirow{2}{*}{\textbf{\makecell[{{p{1.8cm}}}]{\textcolor{blue}{Human (H)}}}} & 
\multirow{2}{*}{\makecell[{{p{6cm}}}]{
The end user who holds intent and externalizes it through text prompts and interactions. Humans do not act on GenAI directly, but influence its behavior only via mediated instructions.}} & 
\textbf{\textcolor{blue}{H}} $\rightarrow$ \textbf{\textcolor{orange}{T}}: The user authors or revises a text prompt to express open-ended goals in natural language. \\&&
\textbf{\textcolor{blue}{H}} $\rightarrow$ \textbf{\textcolor{teal}{I}}: The user performs interactive actions (\eg clicking, highlighting, dragging) on the interface or artifacts to provide information. \\
\hline

\multirow{2}{*}{\textbf{\textcolor{orange}{\makecell[{{p{1.8cm}}}]{Text Prompts (T)}}}} & 
\multirow{2}{*}{\makecell[{{p{6cm}}}]{Natural language instructions written by the user to convey intent. Intuitive but may be ambiguous or incomplete. Used alone or as part of a richer instruction for GenAI}} & 
\textbf{\textcolor{orange}{T}} $\rightarrow$ \textbf{\textcolor{purple}{Aug}}: Text prompts are incorporated into the augmented instruction (combined with interaction-derived information or artifacts). \\&&
\textbf{\textcolor{orange}{T}} $\rightarrow$ \textbf{\textcolor{red}{G}}: The text prompt alone is directly submitted to GenAI as a complete instruction. \\
\hline

\multirow{3}{*}{\makecell[{{p{1.8cm}}}]{\textbf{\textcolor{teal}{Interaction (I)}}}} &
\multirow{3}{*}{\makecell[{{p{6cm}}}]{
Supplemental user actions (\eg clicking, selecting, annotating) that concretize or constrain intent. Interactions do not generate content; they specify how or where intent applies.}} & 
\textbf{\textcolor{teal}{I}} $\rightarrow$ \textbf{\textcolor{purple}{Aug}}: Interaction-derived information is integrated into the augmented instruction. \\&&
\textbf{\textcolor{teal}{I}} $\rightarrow$ \textbf{\textcolor{orange}{T}}: Interactions may modify the text prompt itself. \\&&
\textbf{\textcolor{teal}{I}} $\rightarrow$ \textbf{\textcolor{arti}{A}}: The user's interactions operate on artifacts (\eg highlighting parts of an image) to specify or restrict the target scope. \\
\hline

\textbf{\textcolor{purple}{Augmented Instruction (Aug)}} & 
The combined instruction delivered to GenAI, formed by merging information derived from text prompts, interactions, and artifacts. It encodes the complete intent for the AI and only exists as an input to the GenAI system. & \textbf{\textcolor{purple}{Aug}} $\rightarrow$ \textbf{\textcolor{red}{G}}: The augmented instruction is passed to GenAI for execution. GenAI uses this enriched instruction to generate content or perform actions.\\
\hline

\multirow{2}{*}{\makecell[{{p{1.8cm}}}]{\textbf{\textcolor{arti}{Artifacts (A)}}}} & 
\multirow{2}{*}{\makecell[{{p{6cm}}}]{Domain objects (\eg text, image, code) that serve as both the state users interact with and the possible results of GenAI execution. Artifacts are not instructions.}} & 
\textbf{\textcolor{arti}{A}} $\rightarrow$ \textbf{\textcolor{purple}{Aug}}: User interactions on artifacts (\eg highlighting a paragraph) are incorporated into augmented instruction. \\&&
\textbf{\textcolor{arti}{A}} $\rightarrow$ \textbf{\textcolor{red}{G}}: Artifacts provide contextual input or constraints for GenAI processing. \\
\hline

\multirow{2}{*}{\makecell[{{p{1.8cm}}}]{\textbf{\textcolor{red}{Generative AI (G)}}}} & 
\multirow{2}{*}{\makecell[{{p{6cm}}}]{The model (\eg LLM, diffusion model) that interprets instructions and operate on artifacts. 
Depending on the paradigm, GenAI may act as different roles (\eg generator or clarifier).
}} & 
\textbf{\textcolor{red}{G}} $\rightarrow$ \textbf{\textcolor{arti}{A}}: GenAI generates, transforms, or edits artifacts based on the instruction (\eg creating an image, editing a document). \\&&
\textbf{\textcolor{red}{G}} $\rightarrow$ \textbf{\textcolor{teal}{I}}: GenAI initiates interactions (\eg suggesting follow-up options or UI elements) to solicit further user input. \\
\hline

\end{tabular}
\end{table*}

\subsection{Relations}\label{sec:model_relation}
Relations denote the directed channels through which information, constraints, or control flow between entities. Our modeling goal is to retain only the necessary linkages, ensuring that relations remain semantically meaningful, discriminative, and easy to interpret in a given design. To this end, we adopt three guiding principles. 
\textit{(1) Semantic meaningfulness:} a relation must denote a substantive and interpretable flow of information or control~\cite{Cook2025,Rezwana2023}. For example, representing T~$\rightarrow$~A (a prompt directly manipulating an artifact) is inappropriate, as it obscures the generative process; such flows should instead be realized as T~$\rightarrow$~G~$\rightarrow$~A or T~$\rightarrow$~Aug~$\rightarrow$~G~$\rightarrow$~A.  
\textit{(2) Discriminative value:} a relation should contribute to distinguishing paradigms rather than restating ubiquitous background actions~\cite{Beaudouin-Lafon2004, Liao2023}. For instance, H~$\rightarrow$~A (upload or inspection) is common across nearly all systems and adds little explanatory value, so it is treated as background provisioning rather than a defining relation.  
\textit{(3) Agency and provenance preservation:} a relation must preserve clarity over who initiates and owns an instruction~\cite{Amershi2019, Moruzzi2024}. Thus, flows such as G~$\rightarrow$~T or G~$\rightarrow$~Aug are excluded: while the model may propose candidate prompts or widgets (captured as G~$\rightarrow$~I), they only become active instructions after explicit user or UI mediation (I~$\rightarrow$~T or I~$\rightarrow$~Aug), ensuring agency and traceability are maintained.  

Taken together, these principles justify modeling only a small, purposeful subset of all possible relations rather than the full combinatorial space. For each entity pair (X, Y), we evaluate whether X can produce information or control that Y can meaningfully consume in the context of human–GenAI communication. If so, the relation is included with its semantics recorded; if not, it is excluded with an explicit rationale. \autoref{tab: entity} presents the resulting relation set.  Detailed explanations are as follows: 

\begin{itemize}
\item 
\textbf{Human (H).} Humans possess intent and decision authority but do not execute generation themselves.
Consequently, humans can compose and revise textual instructions (H~$\rightarrow$~T) and perform focused interactions (H~$\rightarrow$~I) such as highlighting, brushing, or selecting. 
Humans also provide or inspect artifacts through interfaces. 
However, since the action of upload or inspection is ubiquitous across systems and does not by itself distinguish paradigms, we treat H~$\rightarrow$~A as background behavior rather than a central comparative relation (no H~$\rightarrow$~A).
Critically, humans do not directly perform generation (no H~$\rightarrow$~G) without going through instructions. 

\item 
\textbf{Text Prompt (T).} 
\review{Q3, Q5.1}\revise{\textbf{T} is a general-purpose, free-form specification of intent in a natural language string, which is manually authored or edited by a human.
As an entity, \textbf{T} represents the text that will be directly sent to GenAI (T~$\rightarrow$~G) or be combined with interaction-derived information to form an augmented instruction (T~$\rightarrow$~Aug). \textbf{T} does not act on artifacts directly (no T~$\rightarrow$~A), and only textual content counts as \textbf{T}; GUI actions or structured editing widgets are not defined as prompts.}

\item 
\textbf{Interaction (I).} 
\review{Q3, Q5.1}\revise{
Interaction (I) refers to non-textual, often GUI user actions (\eg clicking, dragging, brushing, or manipulating structured editors) that can modify, supplement, or operationalize what the text prompt expresses or what the artifact provides, while not themselves constituting valid GenAI inputs.
}
Interactions can operate on artifacts (I~$\rightarrow$~A) like selecting target elements, modifying or augmenting prompts (I~$\rightarrow$~T), or feeding information into the augmented instruction (I~$\rightarrow$~Aug). Interactions cannot evoke GenAI to generate by themselves (no I~$\rightarrow$~G): they are not generators but mediators of specificity. \revise{Furthermore, \textbf{I} excludes the act of writing text prompts as natural language strings without formatting.}


\item 
\textbf{Augmented Instruction (Aug).} Augmented instruction represents the instructions beyond pure NL prompts that the GenAI will execute. 
By definition, Aug is constructed from prompt and interaction inputs (T~$\rightarrow$~Aug, I~$\rightarrow$~Aug) and can additionally incorporate direct artifact-derived context (A~$\rightarrow$~Aug) when a selection or context snippet is encoded into the instruction. 
The only valid execution path from Aug is into the model (Aug~$\rightarrow$~G); Aug does not itself perform artifact edits (no Aug~$\rightarrow$~A). \review{Q3, Q4}\revise{Treating \textbf{Aug} as the explicit input to GenAI makes paradigm comparisons straightforward: different tools differ chiefly in how they build \textbf{Aug} (which combinations of \textbf{T}, \textbf{I}, and \textbf{A} feed into it). 
Please refer to \autoref{sec: atomic paradigms} for more details.
Meanwhile, T~$\rightarrow$~G still remains essential, as users often craft a text prompt before the model initiates further interaction.}

\item 
\textbf{Artifact (A).} 
Artifacts are the domain objects—texts, images, code, datasets—that provide both targets and context. Artifacts are passive in the relation set: they do not autonomously produce text prompts (no A~$\rightarrow$~T) or initiate interactions (no A~$\rightarrow$~I). Relevant relations include I~$\rightarrow$~A (interactions operate on artifacts to select or annotate), A~$\rightarrow$~Aug (artifact content or references can be incorporated into the augmented instruction), and A~$\rightarrow$~G (artifacts can be provided directly as model input). 
The distinction between A~$\rightarrow$~Aug and A~$\rightarrow$~G is meaningful for paradigm design: A~$\rightarrow$~Aug indicates artifact-derived constraints become part of the composite instruction, whereas A~$\rightarrow$~G indicates the artifact (and its extracted features) is supplied as raw model context.

\item 
\textbf{Generative AI (G).} 
GenAI is the executor, it accepts an instruction (T~$\rightarrow$~G or Aug~$\rightarrow$~G) and produces or modifies artifacts (G~$\rightarrow$~A). In mixed-initiative paradigms, GenAI may also present interaction affordances or clarification options (G~$\rightarrow$~I) to solicit further user input. 
GenAI cannot directly produce \textbf{Aug} (no G~$\rightarrow$~Aug), as it must embed user interaction-derived information; likewise, when GenAI generates prompt suggestions, users need to take explicit interaction to turn it into \textbf{T} (\ie no G~$\rightarrow$~T).
We define this constraint following the well-known human-AI interaction guidelines, where users should be aware of AI actions and have full control over AI~\cite{Amershi2019}. 
Without the constraint, AI can prompt itself without human control, violating the principles.
\end{itemize}

The relation set is \textit{necessary} because each retained relation reflects a non-reducible semantic flow, consistent with the principle of semantic meaningfulness. 
It is also \textit{sufficient} because, after systematically enumerating and pruning all possible pairs, the remaining compositions span the full spectrum of prompt–interaction dynamics, aligning with discriminative value and agency preservation. 
We further demonstrate the  descriptive, discriminative, and generative powers of our entity set and relation set through systematically review existing interface design and proposing new variations in \autoref{sec: design paradigm} and \autoref{sec: case}.

\begin{figure*}[t]
  \centering
    \includegraphics[width=0.65\linewidth]{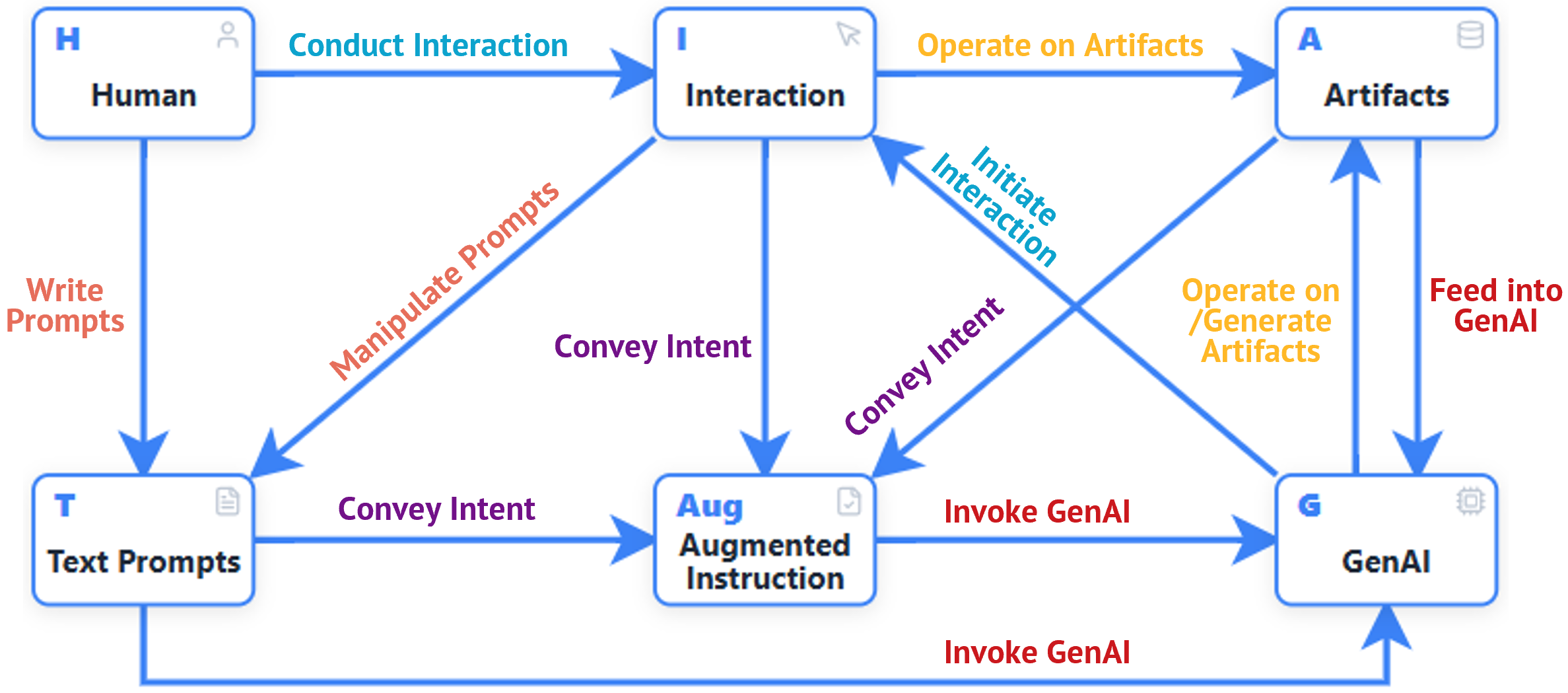}
    \caption{\review{Q7}\revise{Interaction-Augmented Instruction Model. }}
\label{fig: graph}
  \vspace{-10px}
\end{figure*}

\subsection{Atomic Paradigm Graph}\label{sec:model_graph}

\autoref{fig: graph} depicts the full interaction-augmented instruction model as a directed graph. 
However, considering task-specific needs and the complexity of integrating all entities and relations, not all of them must be adopted to represent an individual interface.
Therefore, we introduce the concept of an \textbf{atomic paradigm graph}, defined as a minimal, self-contained subgraph of the model that captures a single, coherent interaction paradigm (\autoref{tab: paradigm}).
Here, atomic means that GenAI (G) plays only one role within the paradigm (avoiding overlapping or conflated functions) and that the paradigm necessarily involves Interaction (I), as paradigms without interactions fall outside our scope.
Each atomic paradigm graph is constructed by sequentially selecting entities and relations in the order they are enacted, thereby encoding the workflow through which a tool supports human–GenAI communication.
Accordingly, a concrete tool can be represented by one or more atomic paradigm graphs, depending on how many distinct paradigms it supports.
For example, an advanced image editing interface with brush-based region selection (\autoref{tab: paradigm}-P4) would add H~$\rightarrow$~I~$\rightarrow$~A~$\rightarrow$~Aug, and route the prompt through T~$\rightarrow$~Aug~$\rightarrow$~G, forming a richer paradigm~\cite{Masson2023b, Liu2024d}. In another variant (\autoref{tab: paradigm}-P8), if GenAI generates widgets to further tune generated artifacts, the graph will include G~$\rightarrow$~I (GenAI initiates widget interactions) followed by H~$\rightarrow$~I~$\rightarrow$~A (human interacts with widgets to act on artifacts)~\cite{Vaithilingam2024}. These small structural differences capture paradigm characteristics.

\section{Design Paradigms}
\label{sec: design paradigm}
To examine if our IAI model can be generally applied to describe and differentiate existing designs, we revisited and annotated 66 GenAI system interfaces that combine prompts and interactions in human-GenAI communications.
We found that all workflows can be represented with atomic graphs derived from the model, demonstrating its descriptive capacity to capture recurring paradigms.
Furthermore, we identified 12 recurring atomic interaction paradigms, which effectively capture similarities and differences between interfaces.
It verifies our IAI model's power of differentiating interaction designs.
We also hope that our identified paradigms can be a starting point for future interface design.


\subsection{Revisit Corpora}\label{sec: corpus}
\review{Q8}\revise{
\nistart{Data}
We revisited prior corpora about interaction-enhanced GenAI interfaces~\cite{Instructions, Gao2024, Cook2025, Luera, Lehmann2024}, which together enumerate 483 candidate instances, and filtered tools according to three criteria: (1) the system involves at least one GenAI model; (2) text prompts are supported for communication; and (3) at least one other interaction modality (\eg selection, brushing, sketching) is used to augment text prompts. 
For each candidate, we examined available demo videos and project pages; otherwise we carefully read the paper to verify compliance. Three authors conducted the initial screening in roughly equal partitions, rotating assignments to ensure every decision was cross-checked. 
Applying these filters yielded 66 representative system interfaces for analysis}.

\nistart{Annotation and Analysis}
Following \autoref{sec:model_graph}, we decomposed each system interface into one or more atomic paradigm graphs (\autoref{fig: graph}) through manual annotation.
During annotation, each atomic graph is instantiated by choosing the entities and relations the system interface implements, and by assigning sequence indices to relations to indicate the temporal or information-flow order. Concurrent flows receive the same index (\eg when a user selection of artifact elements is incorporated into an augmented instruction simultaneously: H~$\rightarrow$~I~$\rightarrow$~A~$\rightarrow$~Aug). 

\review{Q8}\revise{Annotation was conducted by two authors using a structured, multi-stage protocol to ensure reliability and reproducibility. We began with a calibration pilot of 10 randomly sampled systems to test the initial codebook (grounded in IAI model design (\autoref{sec: model}) and definitions (\autoref{tab: entity}), surface ambiguous cases, and refine operational rules.
The two annotators then independently coded half of the remaining systems and cross-checked each other's annotation to reach a shared interpretation grounded in the codebook.
All remaining uncertainties were reviewed and resolved with the involvement of a third author, and the resulting consensus annotations form the basis of our analysis.
In addition to basic model definitions, our annotation followed these principles:


\begin{itemize}
\item \textbf{AI atomicity.} Each atomic paradigm graph assigns GenAI a single, well-defined role to preserve atomicity and avoid ambiguity arising from multiple simultaneous AI functions.
\item \textbf{Interaction requirement.} An atomic paradigm must involve at least one interaction modality; purely prompt-only workflows fall outside our scope.
\item \textbf{Ordered relations.} Relations are numbered to encode the workflow order; relations that occur concurrently share the same index number.
\item \textbf{Iteration elision.} Repetitive iteration (\eg multiple edit cycles with the same interaction design) is not re-annotated, as one representative cycle suffices to capture the paradigm’s structural characteristics.
\end{itemize}

To derive recurring paradigms, we canonicalized tool-specific labels into the standardized entities and relations  and normalized away incidental UI differences that do not affect information flow. We then grouped the resulting atomic graphs by their topological structure to identify recurring interaction patterns. For each topological class, we generated a paradigm name, a canonical graph, a concise description, and a mapping to all systems exhibiting that structure. All tool metadata, annotation data, and grouping information are released to support replication in \url{https://interaction-augmented-instruction.github.io/}.
}

\begin{table*}[t]
\centering
\caption{Taxonomy of atomic interaction paradigms in human-GenAI communication. \re{The tools used to demonstrate corresponding paradigms in \autoref{sec: atomic paradigms} are shown in \textbf{bold}.}}

\label{tab: paradigm}
\renewcommand\arraystretch{1}
\setlength{\tabcolsep}{1mm}{
\resizebox{0.96\textwidth}{!}{%
\begin{tabular}{|m{1.9cm}|m{1.9cm}|m{2.8cm}|m{4.2cm}|m{4.3cm}|m{4.5cm}|}
\hline
  {\centering \textbf{Interaction Timing}} &
  {\centering \textbf{\re{User} \newline Resources}} &
  {\centering \textbf{Paradigm Name}} &
  {\centering \textbf{Paradigm Description}} &
  {\centering \textbf{Atomic Paradigm Graph}} &
    {\centering \textbf{\re{Example Tools}}} \\ 
\hline
  
\multirow{4}{*}{\parbox{1.9cm}{\centering Interaction \textbf{Before} Calling GenAI}}
 &
  \multirow{3}{*}{\parbox{1.9cm}{\centering \textbf{Prompt-only} (no artifact at hand)}} &
    \raggedright \textbf{P1.} Interactive Prompt Enhancement &
  A human selects parts of drafted text prompts for GenAI to refine or expand content. &
  \centering {\includegraphics[width=4.0cm]{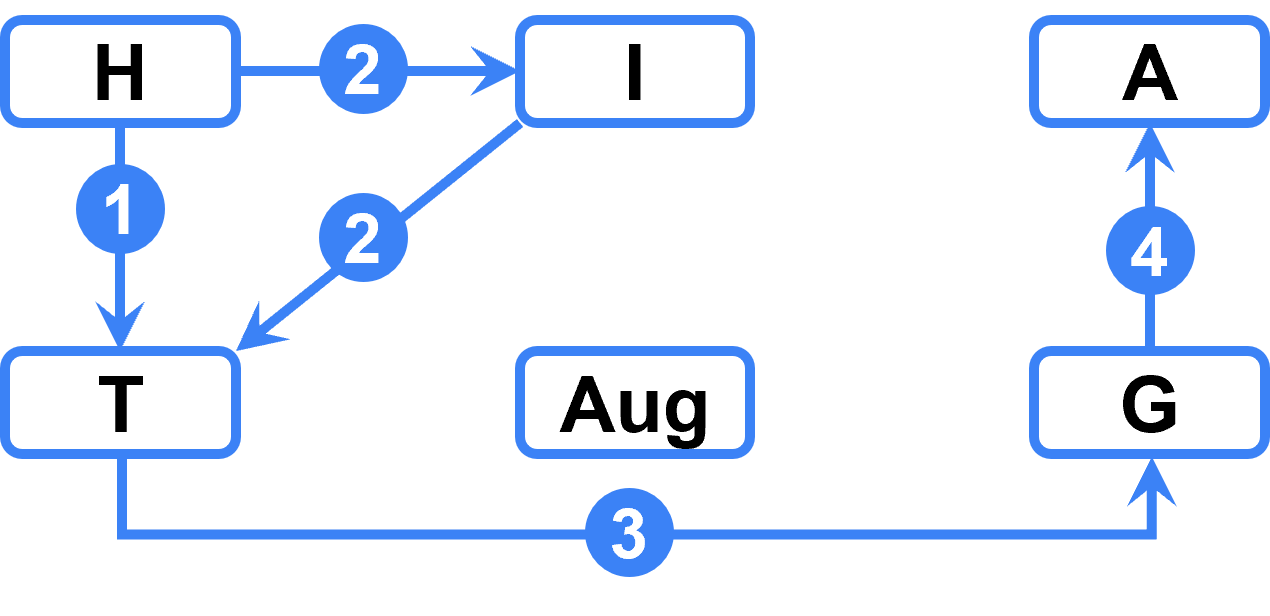}} &
  \textbf{DreamSheets}~\cite{Almeda2024}, Promptify~\cite{Brade2023} \\ \cline{3-6}

 &
   &
  \raggedright \textbf{P2.} Interactive Prompt Organization &
  A human organizes multiple text prompts into structured formats (\eg tree) to fit the application. & \centering{\includegraphics[width=4.0cm]{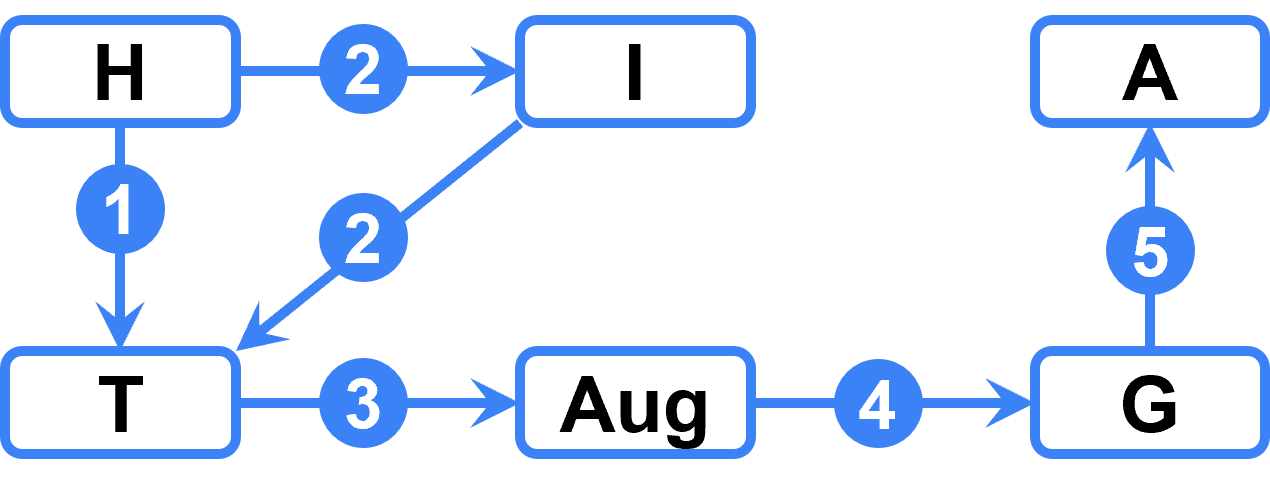}} &
  \textbf{CoLadder}~\cite{Zhu}, CoPrompt~\cite{Feng2024}, PromptChainer~\cite{Wu2022e}, HoT~\cite{Nguyen2023}, IntentTagger~\cite{Gmeiner2025}\\ \cline{3-6}
   &
   &
  \raggedright \textbf{P3.} \review{Q5.4}\revise{Interaction as Part of Instruction} &
  A human’s interactions outside artifacts are included with prompts for GenAI to operate on artifacts. &
  \centering {\includegraphics[width=4.0cm]{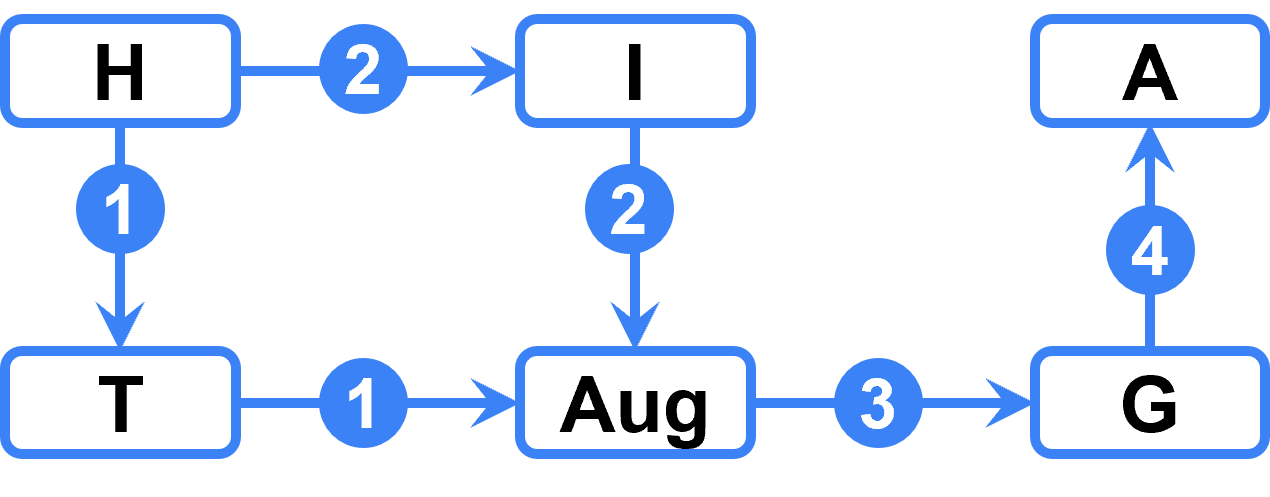}} &
  \textbf{SketchFlex}~\cite{Lin2025a}, InstructPipe~\cite{Zhou2023}, Kim~\etal~\cite{Kim2023d} \\ \cline{2-6}
 &
    \multirow{1}{*}{\parbox{1.9cm}{\centering \textbf{Artifact-grounded}}} &
  \raggedright \textbf{P4.} \review{Q5.4}\revise{Referenced Artifact as Part of Instruction}&
  Human-selected partial or complete artifacts are combined with prompts as GenAI instructions. &
  \centering {\includegraphics[width=4.0cm]{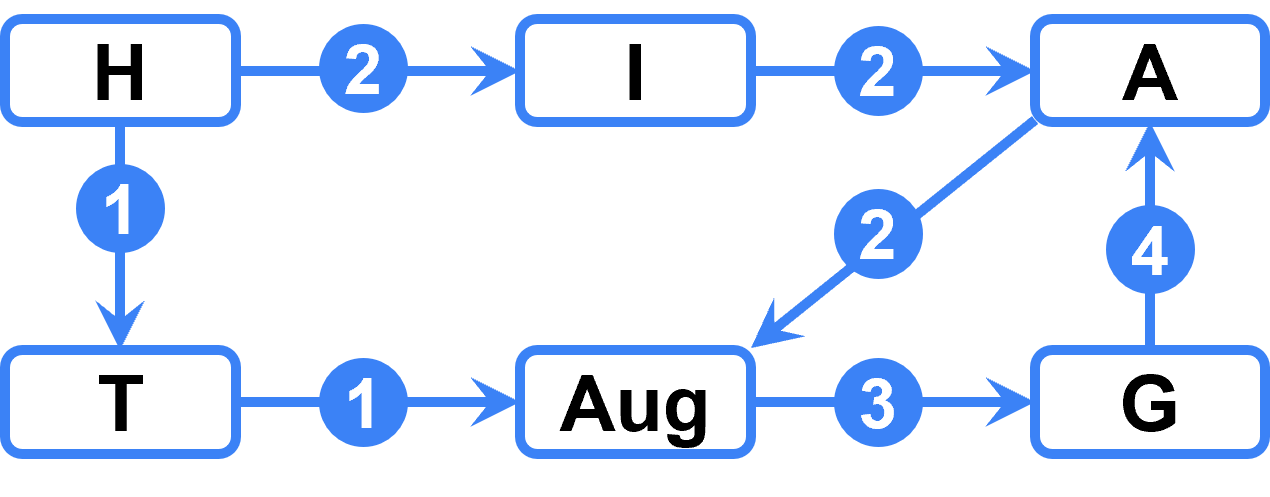}} &
\textbf{DirectGPT}~\cite{Masson2023b}, Data Formulator~\cite{Wang2023a}, ChatScratch~\cite{Chen2024d}, Qlarify~\cite{Fok2024}, InternGPT~\cite{Liu2023c}, MagicQuill~\cite{Liu2024d}, \etc \\ \hline

\multirow{8}{*}{\parbox{1.9cm}{\centering Interaction \textbf{After} \newline Calling GenAI}} &
\multirow{4}{*}{\parbox{1.9cm}{\centering \textbf{Prompt-only} (no artifact at hand)}} &
  \raggedright \textbf{P5.} AI-driven Prompt Suggestion&
  GenAI suggests new or extended prompts from the human’s initial input for selection. &
  \centering {\includegraphics[width=4.0cm]{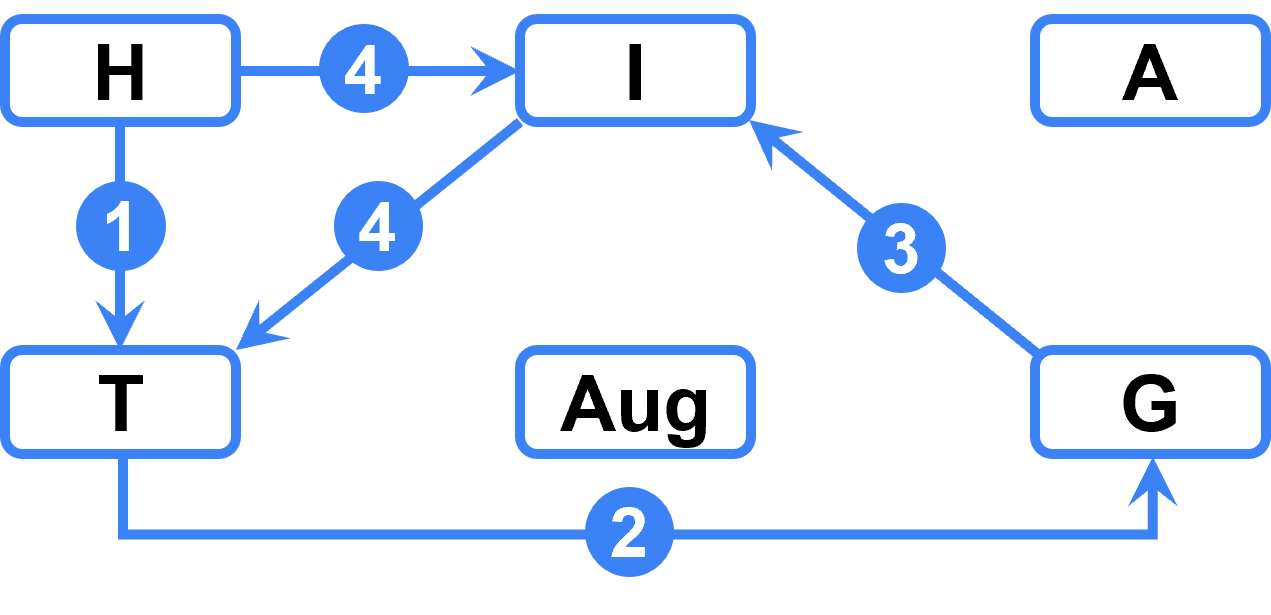}} &
  \textbf{Spellburst}~\cite{Angert2023}, IntentTagger~\cite{Gmeiner2025}, PromptCharm~\cite{Wang2024g}, Bursztyn~\etal~\cite{Bursztyn2021}, PromptCrafter~\cite{Baek2023}, AiCommentator~\cite{Andrews2024}, Sensecape \cite{Suh2023a}, ExploreLLM \cite{Ma2024}, \etc \\ \cline{3-6}
 &
   &
  \raggedright \textbf{P6.} AI-driven Prompt\newline Decomposition &
  GenAI restructures prompts into fine-grained, organized forms for interactive manipulation. &
  \centering {\includegraphics[width=4.0cm]{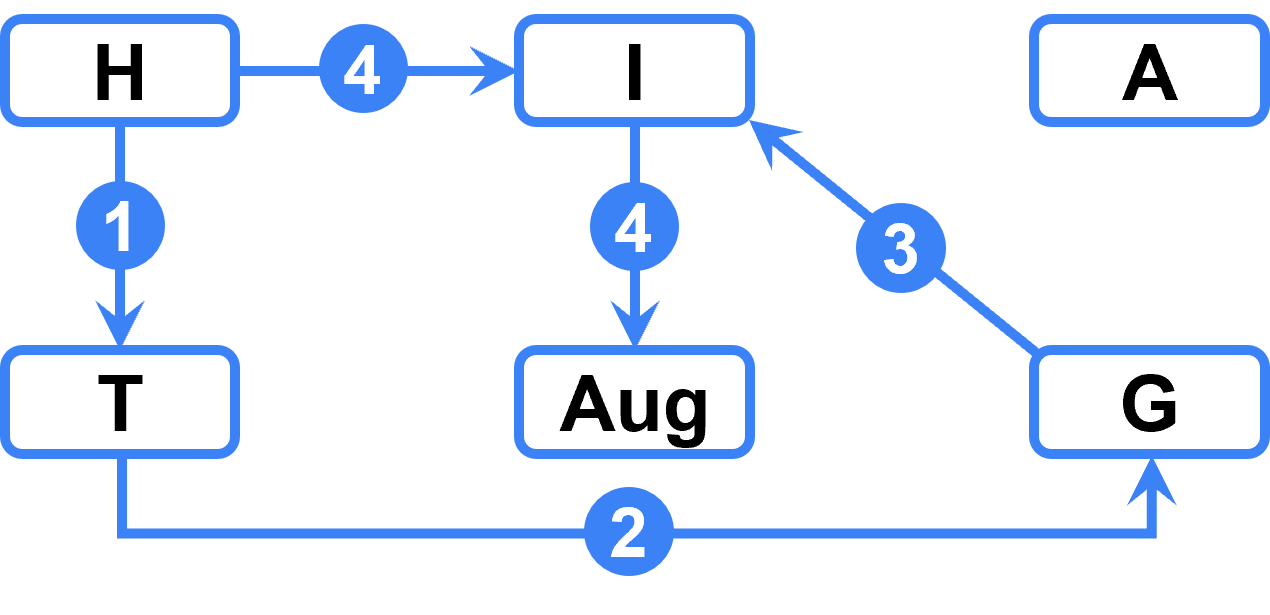}} &
  \textbf{Low-code LLM}~\cite{Cai2024}, NeuroSync~\cite{Zhang2025}, Liu \etal~\cite{Liu2023e}, SimStep~\cite{Kaputa2025}, Fragments~\cite{Riche2025} \\ \cline{3-6}
 &
   &
  \raggedright \textbf{P7.} Generative Prompt Control Widgets &
  GenAI generates interactive widgets for fine-grained prompt control. &
  \centering {\includegraphics[width=4.0cm]{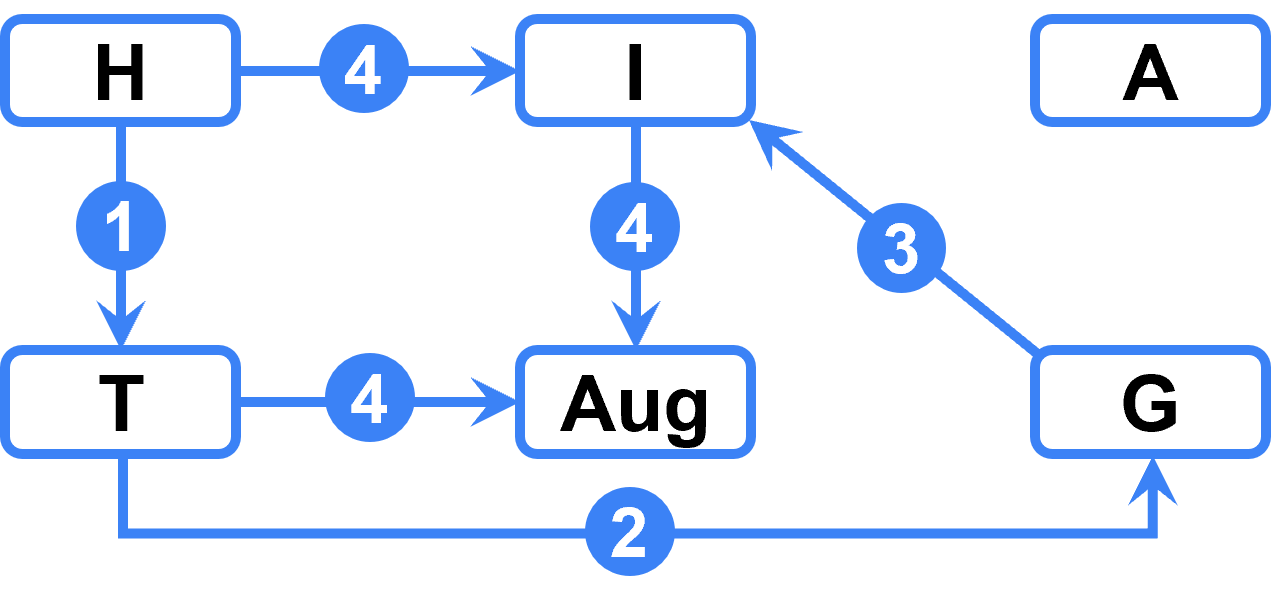}} &
  \textbf{PromptCharm}~\cite{Wang2024g}, DancingBoard~\cite{Chen2025}, IntentTagger~\cite{Gmeiner2025} \\ \cline{3-6}
  &
   &
    \textbf{P8.} Generative Artifact Control Widgets &
  GenAI generates widgets for humans to further manipulate or confirm artifact-related actions. &
  \centering {\includegraphics[width=4.0cm]{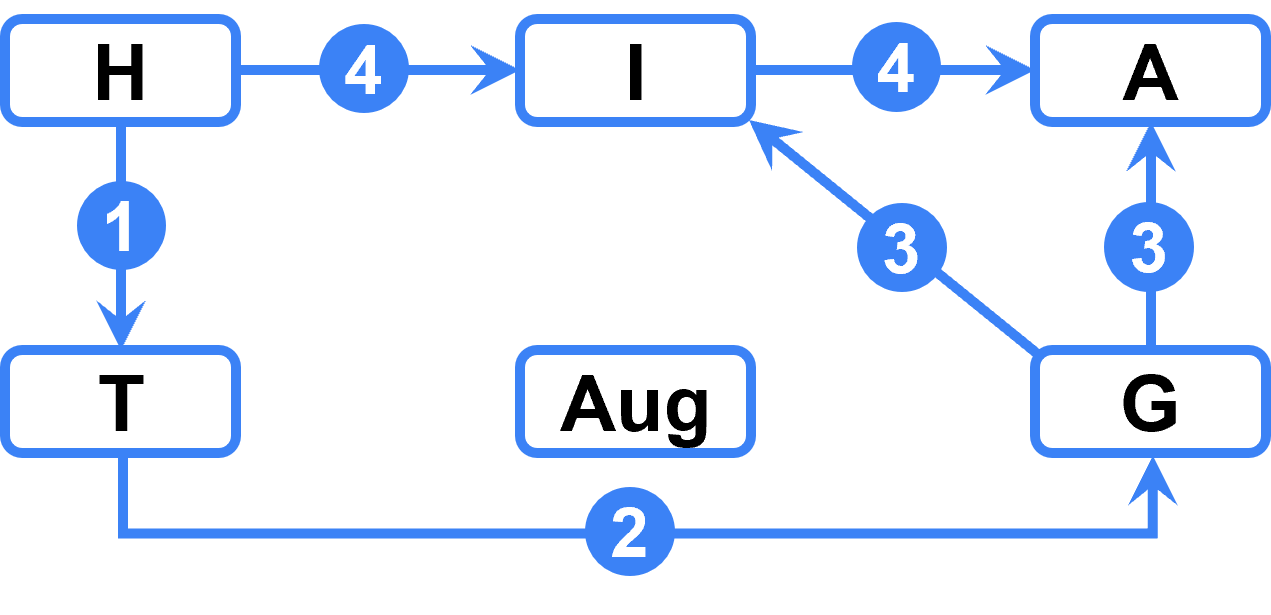}} &
\textbf{DynaVis}~\cite{Vaithilingam2024}, Graphologue~\cite{Jiang2023}, Luminate~\cite{Suh2023}, Sparks~\cite{Gero2022}, ConstitutionMaker~\cite{Petridis2024}, ChatScratch~\cite{Chen2024d}, ScatterShot~\cite{Wu2023c}, \etc\\ \cline{2-6}
 &
\multirow{4}{*}{\parbox{1.9cm}{\centering \textbf{Artifact-grounded}}} &

  \raggedright \textbf{P9.} Artifact to Structured Instruction &
  GenAI uses artifacts as starting points to generate structured instructions. &
  \centering {\includegraphics[width=4.0cm]{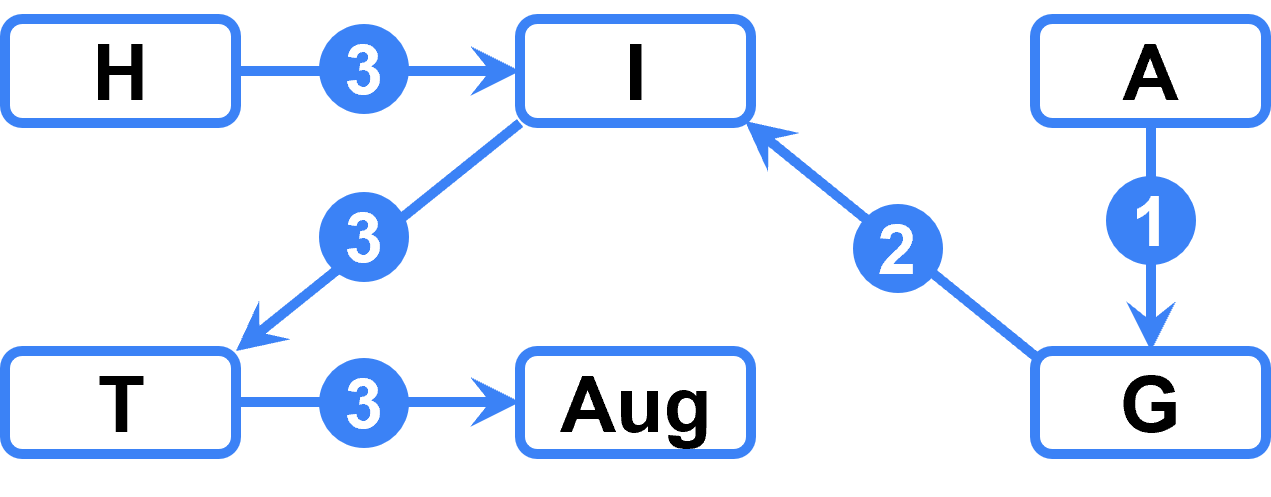}} &
  \textbf{VISAR}~\cite{Zhang2023b}, Metaphorian~\cite{Kim2023c}, IntentTagger~\cite{Gmeiner2025} \\ \cline{3-6}
 &
   &
  \raggedright \textbf{P10.} Artifact to Multimodal Instruction &
  GenAI parses and integrates artifacts to construct multimodal instructions. 
  &
  \centering {\includegraphics[width=4.0cm]{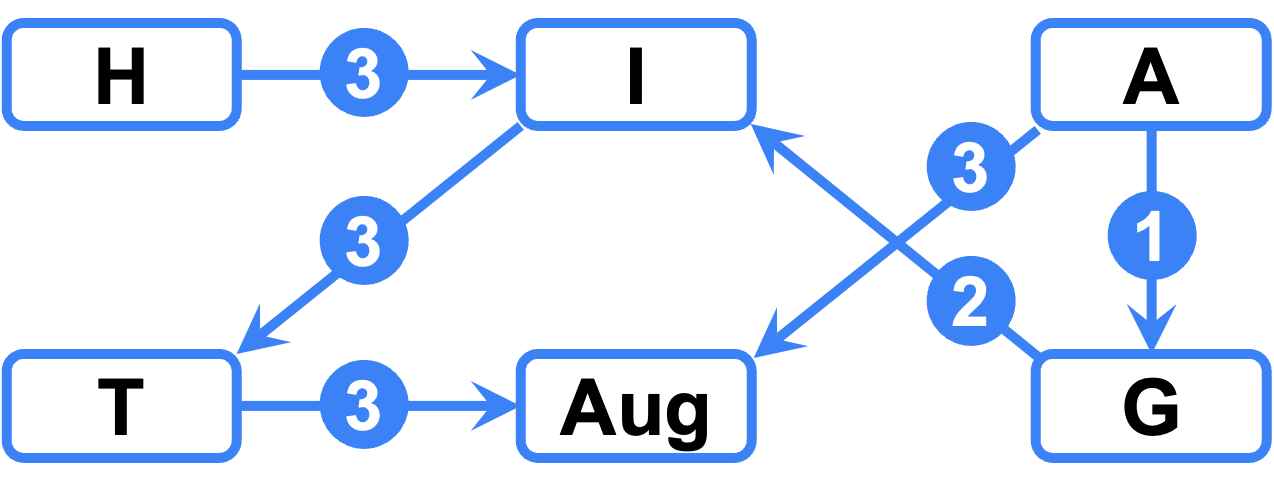}} &
  \textbf{FigurA11y}~\cite{Singh2024} \\ \cline{3-6}
 &
   &
  \raggedright \textbf{P11.} Artifact-driven Prompt Enhancement &
  GenAI suggests actions based on contextual requests integrating artifacts and prompts. &
  \centering {\includegraphics[width=4.0cm]{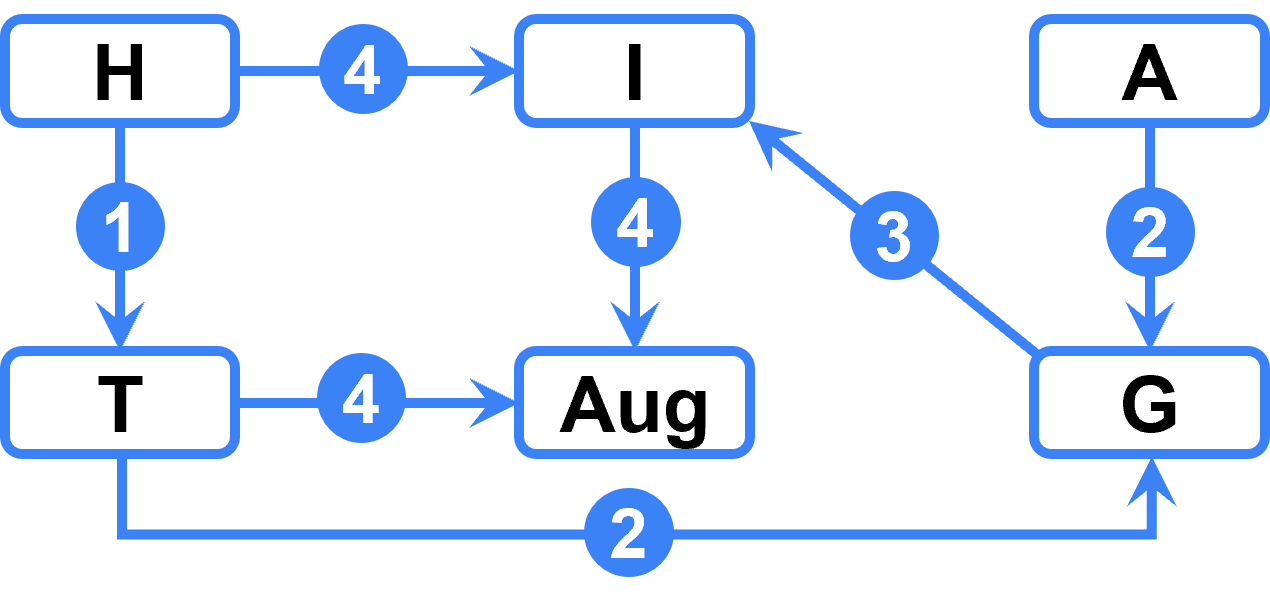}} &
  \textbf{LangAware}~\cite{Chen2023e} \\ \cline{3-6}
 &
   &
     \raggedright \textbf{P12.} Interactive Artifact Refinement &
  GenAI analyzes artifacts and initiates relevant interactions based on prompts. &
  \centering \includegraphics[width=4.0cm]{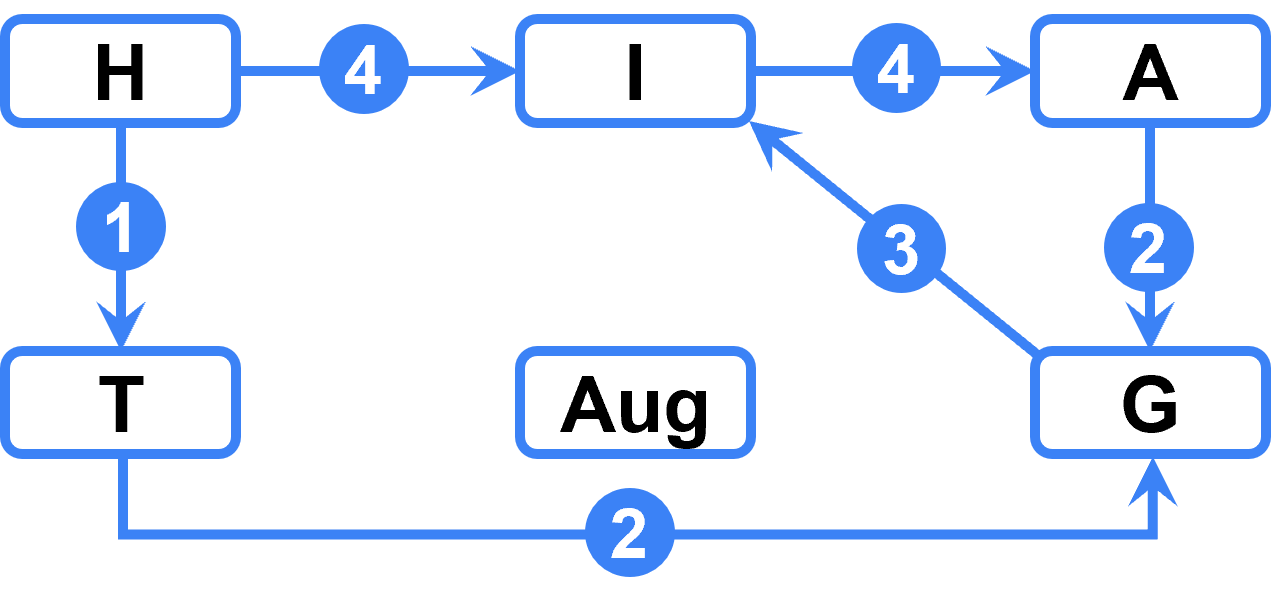} &
  \textbf{PDFChatAnnotator}~\cite{Tang2024} \\ \hline

\end{tabular}%
}}
\vspace{2em}
\end{table*}


\begin{figure*}[t]
  \centering
    \includegraphics[width=\linewidth]{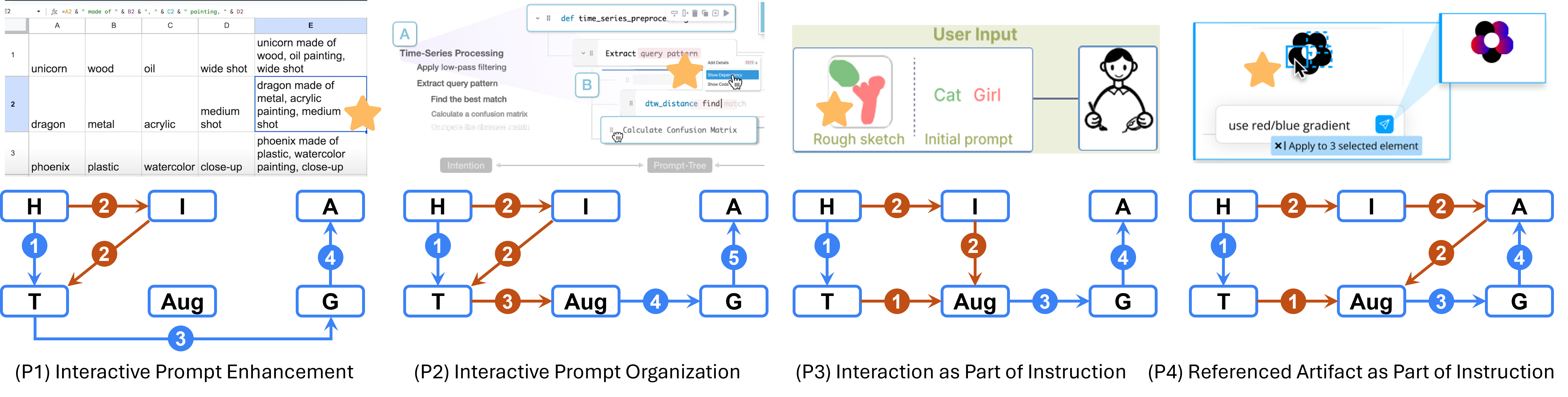}
     \vspace{-10px}
    \caption{
    \review{Q7}\revise{Examples of pre-invocation paradigms \re{(\includegraphics[width=0.017\linewidth]{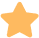} indicates the key interaction design in the interface, and \textcolor[RGB]{192,79,21}{red line} highlights the key differences among the paradigm graphs)}, including prompt-only (P1-P3) and artifact-grounded (P4):
    (P1) Interactive prompt enhancement (\eg Dreamsheets~\cite{Almeda2024}); 
    (P2) Interactive prompt organization (\eg CoLadder~\cite{Zhu});
    (P3) Interaction as part of instruction (\eg SketchFlex~\cite{Lin2025a});
    (P4) Referenced artifact as part of instruction (\eg  DirectGPT~\cite{Masson2023b}). }
    }
\label{fig: paradigm0}
\end{figure*}

\subsection{Atomic Paradigms} 
\label{sec: atomic paradigms}

\begin{figure*}[t]
  \centering
    \includegraphics[width=\linewidth]{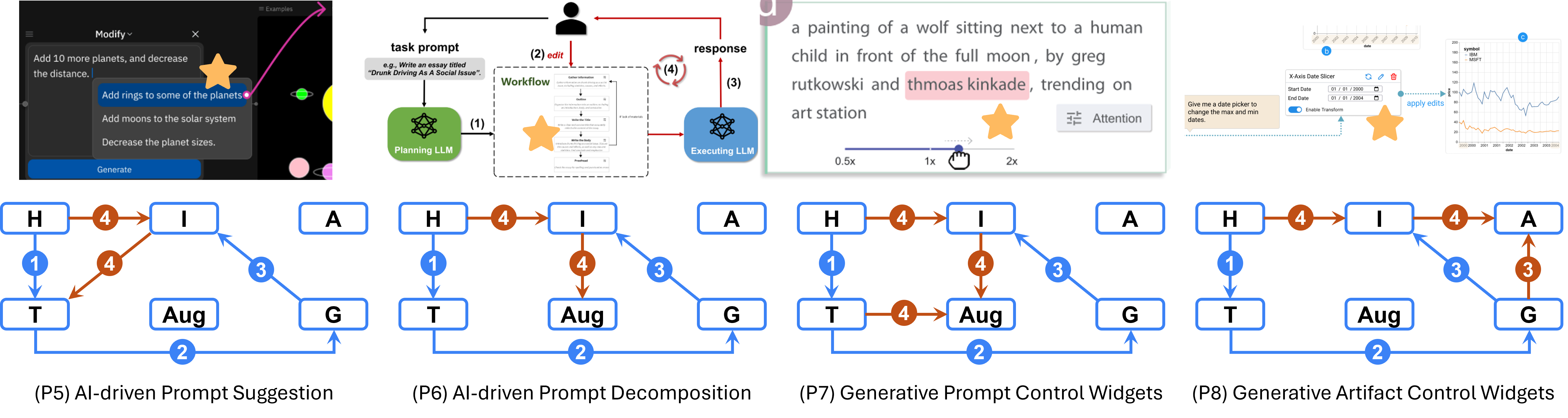}
     \vspace{-10px}
    \caption{
    \review{Q7}\revise{Examples of post-invocation, prompt-only paradigms \re{(\includegraphics[width=0.017\linewidth]{figures/star.png} indicates the key interaction design in the interface, and \textcolor[RGB]{192,79,21}{red line} highlights the key differences among the paradigm graphs)}:
    (P5) AI-driven prompt suggestion (\eg Spellburst~\cite{Angert2023}); 
    (P6) AI-driven prompt decomposition (\eg Low-code LLM~\cite{Cai2024});
    (P7) Generative prompt control widgets (\eg PromptCharm~\cite{Wang2024g});
    (P8) Generative artifact control widgets (\eg DynaVis~\cite{Vaithilingam2024}).}
    }
\label{fig: paradigm1}
\end{figure*}

Using the IAI model as an analytic lens, we identified 12 recurring atomic paradigms (\autoref{tab: paradigm}) from our corpus. We organize these paradigms along two orthogonal dimensions based on our model:
(1) \textbf{interaction timing}: whether Interaction (I) occurs \emph{before} or \emph{after} invoking GenAI (G); and (2) \textbf{user resources}: whether the user begins interacting with GenAI (G) when they have no Artifact (A) at hand (\emph{prompt-only}) or are in the \emph{artifact-grounded} situation. 
The timing dimension roughly tracks intent clarity and control locus: pre-invocation interactions are typical when users can specify constraints up front, whereas post-invocation interactions support exploratory or ambiguous goals via mixed-initiative refinement.
The user resource dimension separates workflows where intent must be expressed solely in natural language from those where an existing artifact can be selected, annotated, or structured to ground the natural language instruction. 
These dimensions yield four paradigm classes with characteristic graph topologies and consistent design applicability.
\review{Q1, Q6}\revise{

\re{
Atomic paradigm graphs are designed to capture human–GenAI interaction logic and information flow, rather than to fully depict end-to-end system execution, and therefore do not necessarily terminate at the same node. For example, in pre-invocation paradigms, GenAI typically acts as a generator: users initiate interactions to construct an augmented instruction, which is then executed by GenAI, making the generated artifact a natural endpoint of the graph. In post-invocation paradigms, by contrast, GenAI often plays a clarifying or interpretive role: it inspects existing artifacts or prior generated results and initiates interactions to elicit user input, usually resulting in a refined instruction rather than a new artifact. Consistent with our annotation principles that each atomic paradigm graph contains only a single GenAI role, passing this refined instruction to a generator GenAI is represented as a separate atomic graph, and complete application workflows can be expressed through chains of such graphs. Paradigms that involve no interaction are outside our scope, as our focus is on characterizing interaction structures and information flow rather than all possible execution paths.}

To better showcase the model’s descriptive and discriminative power, accordingly, the following subsections adopt a unified structure: we first describe the shared characteristics and typical cases of each paradigm class; we then analyze how similarities and differences in their paradigm graph structures map to concrete operations of each tool through representative comparisons (\autoref{fig: paradigm0}, \autoref{fig: paradigm1}, and \autoref{fig: paradigm2}); 
and we conclude each subsection by articulating the key intra-class distinctions and the design implications they entail.}


\subsubsection{Pre-invocation, Prompt-only: Structuring and Enhancing Prompts Before GenAI Invocation}
\label{sec: pre-call prompt}
This class groups paradigms (\autoref{tab: paradigm} P1-P3) where users begin with only prompts and seek to specify intent through pre-invocation refinement. 
Users typically have a clear goal and employ interaction to organize, extend, or transform textual instructions so that the ensuing generation aligns with their intent more precisely. For instance, a writer may elaborate a draft prompt with inline selection~\cite{canvas}, or a programmer may decompose the prompt for a coding task into structured prompts for subtasks~\cite{Feng2024}.

In all three paradigms (\autoref{fig: paradigm0}~P1-P3), humans write prompts (H~$\rightarrow$~T) and GenAI produces artifacts (G~$\rightarrow$~A), yet they diverge in how to enhance the original natural language prompts (\ie refine \textbf{T} or construct \textbf{Aug}).
The first paradigm, \textit{interactive prompt enhancement} (P1), keeps the representation within the boundaries of natural language. User interactions just refine the wording of \textbf{T} (H~$\rightarrow$~I~$\rightarrow$~T), which is then executed directly. 
Dreamsheets~\cite{Almeda2024}, for instance, enables users to rapidly compose text prompt variations with a set of keywords through spreadsheet-like interactions for exploratory image generation \review{Q1}\revise{(\autoref{fig: paradigm0}-P1 and \autoref{fig:motivating-case}-B)}.
While effective for quick iteration, it remains limited to text and cannot capture richer logical structures.
In \textit{interactive prompt organization} (P2), interactions introduce additional structure beyond natural language. The enriched representation, encoded as augmented instruction, integrates hierarchical or compositional logic before being passed to GenAI (H~$\rightarrow$~I~$\rightarrow$~T~$\rightarrow$~Aug). CoLadder~\cite{Zhu} illustrates this by arranging prompts into a tree of subtasks (\autoref{fig: paradigm0}-P2), while PromptChainer~\cite{Wu2022e} links prompts in sequential chains.
Yet even structured text has some limits; it presumes all user intent can be expressed linguistically. This gap is addressed by the third paradigm.
\textit{Interaction as part of instruction} (P3) encodes user actions themselves as operative intent, introducing non-linguistic signals in addition to text (H~$\rightarrow$~I~$\rightarrow$~Aug; T~$\rightarrow$~Aug). SketchFlex~\cite{Lin2025a}, for example, interprets freehand sketches on a canvas as executable instructions combined with prompts to guide expressive image generation \review{Q1}\revise{(\autoref{fig: paradigm0}-P3 and \autoref{fig:motivating-case}-C)}.

\review{Q6}\revise{The key distinction lies in whether interaction only edits text prompts, introduces structured logic into augmented instruction, or adds non-linguistic information into augmented instruction. The interactive prompt enhancement paradigm (P1) emphasizes speed and lightweight iteration, while the prompt organization paradigm (P2) enables intent decomposition and traceability. Interaction-as-instruction (P3) broadens expressivity by moving beyond natural language altogether.}

\subsubsection{Pre-invocation, Artifact-grounded: Grounding Instructions in Existing Artifacts Before GenAI Invocation}
\label{sec: pre-call artifact}
This class  (\autoref{tab: paradigm} P4) captures paradigms where users start with an artifact and specify intent through pre-invocation manipulation (\eg selection, brushing, sketching).
It is also among the most common paradigms in existing systems.
Typical scenarios include selecting a chart element to query~\cite{Masson2023b}, brushing an image region for editing~\cite{Liu2024d}, or highlighting code for debugging~\cite{cursor}. Here, interactions ground the prompt in concrete referents before generation.

\review{Q7}\revise{In the \textit{referenced artifact as part of instruction} paradigm (P4), humans interact with artifacts to encapsulate either partial or entire content into augmented instruction} (H~$\rightarrow$~I~$\rightarrow$~A~$\rightarrow$~Aug; T~$\rightarrow$~Aug). Systems such as DirectGPT~\cite{Masson2023b} and MagicQuill~\cite{Liu2024d} demonstrate this approach by passing dragged or brushed elements as precise constraints \review{Q1}\revise{(\autoref{fig: paradigm0}-P4 and \autoref{fig:motivating-case}-D)}. 
At first glance, DirectGPT might look similar to SketchFlex
\review{Q1}\revise{(\autoref{fig: paradigm0}-P3 and \autoref{fig:motivating-case}-C)}, as they both allow humans to create visual augmented instruction with interactions for GenAI to operate on artifacts.
However, with our paradigm graphs, it is easy to notice that the key difference lies in how augmented instruction is composed: DirectGPT, as a representative of the \textit{referenced artifact as part of instruction} paradigm, includes raw artifact segments, while SketchFlex, which belongs to the \textit{interaction as part of instruction} paradigm category, encodes non-linguistic signals with interactions (H~$\rightarrow$~I~$\rightarrow$~Aug~$\rightarrow$~G~$\rightarrow$~A).
This comparison highlights the key difference between the two paradigms: whether the interaction carries intent solely without embedding artifact.

\review{Q6}\revise{The design difference shapes how intent ambiguity or unclarity is resolved. The \textit{referenced artifact as part of instruction} paradigm (P4) mitigates referential ambiguity by passing raw content to GenAI, whereas the \textit{interaction as part of instruction} paradigm (P3) reduces descriptive effort by letting interactions themselves encode intent. For editing and debugging tasks, pre-invocation artifact-grounded interactions are particularly effective as they tie model operations to specific referents.}

\subsubsection{Post-invocation, Prompt-only: Iterative Prompt Steering After GenAI Invocation}
\label{sec: post-call prompt}
This class  (\autoref{tab: paradigm} P5-P8) captures cases where users specify intent through prompts and perform interactions to clarify intent \emph{after} an initial GenAI invocation. 
Unlike pre-invocation paradigms, 
paradigms in this class often leverage GenAI for prompt steering and iterative negotiation of intent rather than directly instructing them for final artifact creation.
This class arises when users’ initial instructions are vague, exploratory, or underspecified, and GenAI takes an active role in shaping subsequent prompts. Rather than users knowing exactly what they want, the system helps steer the process step by step.


The first three paradigms in this class (\autoref{fig: paradigm1} P5-P7) share the basic flow H~$\rightarrow$~T~$\rightarrow$~G (human writes text prompts to GenAI) and G~$\rightarrow$~I (GenAI initiates interactions) but diverge in how AI-initiated interactions reshape subsequent instructions.
In \textit{AI-driven prompt suggestion} (P5), GenAI generates candidate refinements or alternative text prompts that users can adopt or edit (H~$\rightarrow$~I~$\rightarrow$~T), focusing on direct modification of natural language. Spellburst~\cite{Angert2023}, for example, generates multiple auto-completed prompt suggestions by AI models.
Then users can select one to guide subsequent calls (\autoref{fig: paradigm1}-P5).  
Similar to the extension from P1 to P2 (see \autoref{sec: pre-call prompt}), simple prompt refinement or suggestion by AI can hardly facilitate the needs for more structured intent communication.
In \textit{AI-driven prompt decomposition} (P6), GenAI externalizes its internal interpretation of a high-level prompt into a structured form such as a graph or task tree. Users then manipulate this representation directly (H~$\rightarrow$~I~$\rightarrow$~Aug), fragmenting, reordering, or parameterizing subtasks. Low-code LLM~\cite{Cai2024} and a recent work NeuroSync~\cite{Zhang2025}, exemplify this approach by visualizing inferred code-generation plans as editable graphs (\autoref{fig: paradigm1}-P6).
These paradigms still operate primarily through text manipulation, but finer-grained specifications remain difficult to express.

Furthermore, \textit{generative prompt control widgets} (P7) extend prompts beyond natural language by having GenAI synthesize interactive controls. PromptCharm~\cite{Wang2024g}, for instance, generates sliders tied to text spans, enabling users to adjust their relative weights within a text-to-image prompt \review{Q1}\revise{(\autoref{fig: paradigm0}-P7 and \autoref{fig:motivating-case}-E)}.
A related but more hybrid paradigm is \textit{generative artifact control widgets} (P8), where GenAI produces both an artifact and associated widgets that persist for subsequent manipulation. DynaVis~\cite{Vaithilingam2024}, for example, augments visualization-oriented natural language interfaces with dynamic controls that let users iteratively adjust and replay edits with instant feedback (\autoref{fig: paradigm1}-P8).
While P7 and P8 both involve model-generated widgets, their scope differs: P7 parameterizes \emph{prompts}, extending intent specification, whereas P8 binds controls to \emph{artifacts}, turning outputs into malleable, persistent interaction surfaces. This distinction demonstrates the discriminative power of the IAI model and connects directly to recent work on malleable UIs and GenUI~\cite{Min2025a}, suggesting a trajectory where GenAI acts not only as a content generator but also as a co-designer of interfaces through which users iteratively shape intent.

\begin{figure*}[t]
  \centering
    \includegraphics[width=\linewidth]{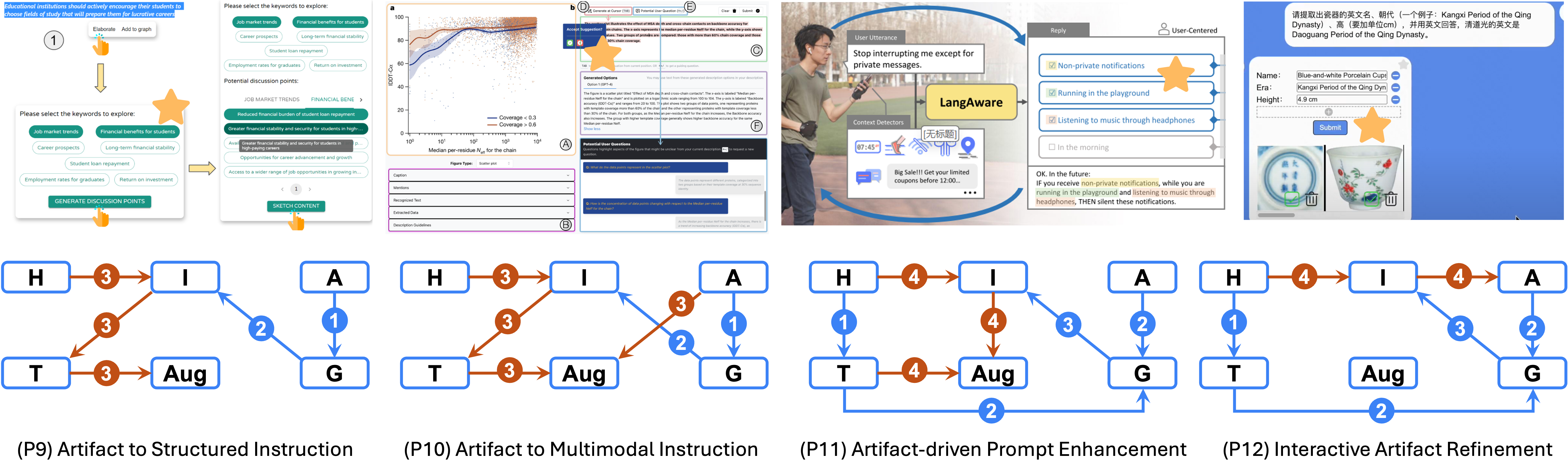}
     \vspace{-10px}
    \caption{
    \review{Q7}\revise{Examples of post-invocation, artifact-grounded paradigms \re{(\includegraphics[width=0.017\linewidth]{figures/star.png} indicates the key interaction design in the interface, and \textcolor[RGB]{192,79,21}{red line} highlights the key differences among the paradigm graphs)}:
    (P9) Artifact to structured instruction (\eg VISAR~\cite{Zhang2023b}); 
    (P10) Artifact to multimodal instruction (\eg FigurA11y~\cite{Singh2024});
    (P11) Artifact-driven prompt enhancement (\eg LangAware~\cite{Chen2023e});
    (P12) Interactive artifact refinement (\eg PDFChatAnnotator~\cite{Tang2024}).}
    }
\label{fig: paradigm2}
\end{figure*}

\review{Q6}\revise{These paradigms differ in where GenAI inserts initiative and how its proposals feed back into instruction. 
AI-driven prompt suggestion (P5) preserves the simplicity of textual prompts but accelerates exploration; 
decomposition (P6) reveals latent reasoning as manipulable structures, enhancing transparency and task management; 
prompt control widgets (P7) enrich prompt expressivity by exposing hidden parameters for direct control; 
and artifact control widgets (P8) extend this idea further by binding widgets to concrete outputs, enabling users to iteratively modify artifacts through persistent controls. }

\subsubsection{Post-invocation, Artifact-grounded: Interactive Editing and Clarification on Artifacts After GenAI Invocation}
\label{sec: post-call artifact}
This class (\autoref{tab: paradigm}, P9–P12) captures workflows in which an artifact serves as the anchor for subsequent interaction, regardless of whether the artifact is produced by a prior GenAI call or provided directly by the user.
Rather than relying solely on textual prompts, GenAI inspects the artifact, initiates interactions, and elicits user feedback to clarify or refine intent. Typical tasks include iteratively editing an image, interrogating a visualization, or correcting model-produced code. 

The common structure is artifact-driven (A~$\rightarrow$~G), with GenAI initiating interactions (G~$\rightarrow$~I) that engage human responses (H~$\rightarrow$~I). The four paradigms differ in how these interactions shape the construction of follow-up instructions.
In \textit{artifact to structured instruction} (P9), GenAI analyzes the artifact, generates a set of candidate operations or reformulations, and presents choices that users select to compose a structured textual prompt. For example, VISAR~\cite{Zhang2023b}  proposes expansion options for a selected paragraph; the user's choices produce a structured prompt tree for subsequent text generation \review{Q1}\revise{(\autoref{fig: paradigm0}-P9 and \autoref{fig:motivating-case}-F)}.
Building on this, \textit{artifact to multimodal instruction} (P10) enriches the workflow by incorporating artifact itself. Here, GenAI extracts salient features from the artifact and asks users to tag or select them; the results are encoded into multimodal augmented instruction that combines artifact content with user intent as prompt. FigurA11y~\cite{Singh2024}, for instance, extracts figure components and lets users link them to accessibility guidelines, producing a multimodal instruction (\autoref{fig: paradigm2}-P10). 
Moving beyond selection, \textit{artifact-driven prompt enhancement} (P11) often positions the artifact as contextual grounding for semantic refinement. GenAI proposes contextual rules or mappings, which users confirm or modify in situ with interactions. LangAware~\cite{Chen2023e} illustrates this by connecting low-level sensor signals to high-level contexts and enabling users to interactively combine contextual information with the original prompt before final execution (\autoref{fig: paradigm2}-P11). 
Finally, \textit{interactive artifact refinement} (P12) emphasizes iterative analysis and improvement. GenAI inspects the artifact in response to a prompt, surfaces candidate elements or annotations, and invites users to inspect, correct, or refine them. PDFChatAnnotator~\cite{Tang2024}, for example, extracts information from PDFs and lets users guide annotation corrections interactively (\autoref{fig: paradigm2}-P12). Unlike prior paradigms, this focuses less on instruction construction and more on artifact-centered troubleshooting and refinement.


\review{Q6}\revise{These paradigms address a recurring user need when users have existing artifacts: in the follow-up communication with GenAI, it might be cumbersome to fully write prompts by themselves (in P9 and P10) or to specify where or how operations should apply precisely (in P11 and P12).
By leveraging GenAI for artifact resolution, these paradigms introduce mechanisms for grounded instruction generation or understanding.
With these paradigms, users can confirm or change with interactions conveniently. }

\begin{table*}[t]
\centering
\small
\caption{\review{Q6, Q10}\revise{Operationalizing Descriptive and Discriminative Power: Design cheat sheet mapping common design situations to recommended paradigms.}
}
\label{tab:cheatsheet}
\renewcommand\arraystretch{1.2}
\setlength{\tabcolsep}{0.8mm}{
\resizebox{\textwidth}{!}{%
\begin{tabular}{m{6.9cm}m{10.6cm}>{\raggedleft\arraybackslash}m{1.1cm}}
\toprule
\textbf{Design Situation} & \textbf{Design Rationale Based on Our IAI Model} & \textbf{Paradigm} \\
\midrule

Need fast, lightweight iterations on text prompts &
Interaction only edits text (H→I→T), keeping the instruction purely textual. &
P1 \\

Need structured intent or task decomposition &
Interaction adds an explicit structure to text (H→I→T→Aug). &
P2 \\

Need non-linguistic or spatial expressivity &
Interaction supplements the instruction (H→I→Aug, T→Aug) beyond text. &
P3 \\\hline

Need to reduce referential ambiguity via artifacts &
Artifact content grounds intent (I→A→Aug, T→Aug) for precise reference. &
P4 \\\hline

Need GenAI-assisted exploration after an initial output &
GenAI proposes refinements (G→I); users adjust via light edits (H→I→T). &
P5 \\

Need transparent intermediate reasoning &
GenAI exposes editable decomposition (G→I) for human adjustment (H→I→Aug). &
P6 \\

Need parameter-level control without editing text prompts &
GenAI initiates parameter widgets (G→I) that augment prompts (H→I→Aug, T→Aug). &
P7 \\

Need persistent, artifact-bound controls for iteration &
Generated widgets (G→I) are attached to artifacts, enabling repeated edits (H→I→A). &
P8 \\\hline

Need to distill follow-up instructions from existing artifacts &
GenAI analyzes the artifact to propose structured follow-ups (A→G→I, H→I→T→Aug). &
P9 \\

Need to form instructions with existing artifacts as references &
GenAI derives multimodal instruction candidates that incorporate artifacts (A→Aug). &
P10 \\

Need artifact-derived suggestions to improve prompts&
GenAI proposes artifact-based actions (A→G→I) that supply prompts (H→I→Aug, T→Aug). &
P11 \\

Need explicit control over how operations apply to artifacts &
GenAI interprets artifacts to initiate grounded interactions (A→G→I) on artifacts (H→I→A). &
P12 \\

\bottomrule
\end{tabular}}}
\end{table*}

\subsection{Cross-paradigm Insights and Suggestions}
\label{sec:cross_insights}


In the previous section, we illustrate paradigm details and discuss their distinct design affordances based on recurring graph topologies derived from our IAI model.
These paradigms and related findings demonstrate our model's meaningful \textit{descriptive} and \textit{discriminative} powers to distinguish interface designs that prior frameworks conflate.
For example, the motivating cases (\autoref{fig:motivating-case} B–F) are clearly mapped to distinct paradigms (P1, P3, P4, P7, P9).
We consider one critical resource of the powers is the IAI model’s information-flow structure.
The structure allows us to use interaction timing and artifact availability as two principal axes to differentiate paradigms.

Starting from interaction timing (C1) and artifact availability (C2), this section
provides concrete, model-grounded considerations that help designers decide which paradigm class aligns with design situations, as shown in \autoref{tab:cheatsheet} (C3), and how the IAI model and derived paradigms can be used to \textit{generate} new interfaces (C4).

\review{Q6}\revise{\nistart{C1. Use interaction timing (interaction before/after invoking GenAI) to choose a paradigm class}}
Interaction timing indexes whether a task’s intent is knowable up front (pre-invocation) or emerges through exploration (post-invocation).
Pre-invocation paradigms assume the user can articulate explicit and clear intent in advance for generating or manipulating artifacts \review{Q6}\revise{(\autoref{tab:cheatsheet} P1-P4)}, such as clear coding structure and logic~\cite{Feng2024}, or explicit image editing areas and tasks ~\cite{Liu2024d}. 
Post-invocation paradigms assume underspecified or exploratory goals \review{Q6}\revise{(\autoref{tab:cheatsheet} P5-P12)}: the system first initiates interactions to further clarify user intents~\cite{Wang2024g} or produces outputs for steering and refinement~\cite{Zhang2025}. \review{Q10}\revise{Designers should align timing with the task's nature based on the model's discriminative power: employ pre-invocation paradigms for tasks requiring precision and auditable outcomes, but prioritize lightweight post-invocation paradigms when the goal is creative ideation and iterative refinement of an ambiguous intent.}

\review{Q6}\revise{\nistart{C2. Use artifact availability (prompt-only vs. artifact-grounded) to select how intent should be expressed}}
Instructions in natural language alone can be ambiguous (\eg ``\textit{make the flower brighter}'' without specifying which flower). The presence of an artifact shifts intent expression from purely linguistic descriptions to concrete, grounded references \review{Q6}\revise{(\autoref{tab:cheatsheet}-P4, P9-P12)}.
By enabling users to select, highlight, or annotate artifact fragments, systems can reduce ambiguity and provide a stable foundation for commands~\cite{Singh2024, Chen2023e}.
The consideration also aligns with design principles for AI-instruments by Riche~\etal~\cite{Riche2025}, where they proposed that text prompts should be grounded in other artifacts.
\review{Q10}\revise{Designers should identify whether their system will receive an initial artifact; if so, artifact-grounded paradigms offer lower ambiguity, richer referential precision, and reduced prompting burden.}

\review{Q6}\revise{\nistart{C3. Within a chosen class, refine toward a specific paradigm based on how intent should be encoded}}
In these design paradigms, a key design choice is how user intent is materialized to augment the original text prompt.
The first way is to directly edit or extend the text prompt with interactions (\autoref{tab: paradigm}-P1, P5).
An example is DreamSheets~\cite{Almeda2024}, where users can combine prompt fragments easily with interactions.
It provides a simple and direct way to improve original prompts but lacks the power to cater complex intent, such as non-linguistic or referential intent.
To handle more complex intent, an approach is to leverage interactions to introduce additional non-linguistic information to text and form augmented instructions, such as structural or parametric information (\autoref{tab: paradigm}-P2, P3, P6, P7, P9 and P11).
For example, CoPrompt~\cite{Feng2024} allows users to drag and drop prompt fragments to form a multi-level list of instructions.
It materializes the non-linguistic intent expression in a convenient way with suitable interaction design.
Regarding the referential intent, the key interaction is to link text prompts with artifacts and generate augmented instructions (\autoref{tab: paradigm}-P4, P10).
It addresses the challenges in describing the linkage between texts and artifacts by materializing users' intent with direct interactions on artifacts.
Lastly, a unique case is to generate new interaction widget by text prompts and facilitate intent expression through direct manipulation (\autoref{tab: paradigm}-P8, P12).
GenAI creates reusable and persistent interactive widgets to materialize the meta intent and allows follow-up similar intent expression with simple interactions with widgets.
A notable example is DynaVis~\cite{Vaithilingam2024}.
The meta intent like changing visual element colors can be materialized as a widget.
The specific color to use can be directly selected by interactions.
\review{Q10}\revise{The summarized design cheat sheet mapping common design situations to recommended paradigms is shown in \autoref{tab:cheatsheet}. These differences correspond to distinct subgraph configurations in the atomic paradigm graphs. Designers can therefore match their intent-expression needs (\eg decomposition, parameterization, selection specificity) to the paradigm whose graph topology supports that form of control.}
\begin{table*}[t]
\small
\caption{\review{Q9}\revise{Operationalizing Generative Power: Potential methods for designers and researchers to apply the IAI model and derived paradigms. In this table, entry points refer to whether the designer or researcher should look into IAI model or our summarized paradigms at the beginning of the usage method.}}
\label{tab:design-guidelines}
\centering
 \vspace{-5px}
\renewcommand\arraystretch{0.9}
\setlength{\tabcolsep}{1.1mm}{
\resizebox{\textwidth}{!}{%
\begin{tabular}{p{3.4cm} p{1.8cm} p{9.6cm}}
\toprule
\textbf{Design Goal} & \textbf{\re{Entry  Point}} & \textbf{Core Workflow} \\
\midrule
\textbf{Extend the Pipeline of an Existing Tool} & \textit{Paradigm} &
1. Identify the atomic paradigm of the existing tool, analyze the paradigm graph, and identify limitations. \\
(refer to \autoref{sec: case 1})& & 2. \textit{When noticing the needs for an additional invocation of GenAI} (\eg asking additional questions for intent clarification), select and chain a complementary paradigm. \\
& & 3. Verify the chained paradigm and introduce new designs. \\
\cmidrule{1-3}
\textbf{Refine the Interaction Design of an Existing Tool} & \textit{Paradigm} &
1. Identify the atomic paradigm of the existing tool, analyze the paradigm graph, and identify limitations. \\
(refer to \autoref{sec: case 2})& & 2. \textit{When it is not necessary for an additional invocation of GenAI} (\eg adding artifacts as part of intent for accurate intent expression), add or adjust targeted relations. \\ 
& & 3. Reinterpret the new graph and introduce new designs. \\
\cmidrule{1-3}
\textbf{Design a New Tool for an Emerging Scenario} & \textit{Model} &
1. Identify IAI model entities in the new scenario by considering scenario requirements and user resources. \\
(refer to \autoref{sec: case 3})
& & 2. Derive the relations based on interaction logic about timing and user resources. \\
& & 3. Map to existing paradigms, get inspiration from example tools in other scenarios, and design the new tool. \\

\cmidrule{1-3}
\textbf{Hypothesize and Research Novel Paradigms} & \textit{Paradigm/Model} & 
1. Select base paradigms from existing ones or design base paradigms from scratch with the IAI model. \\
(refer to \autoref{sec: case 4})& & 2. Experiment inverting, changing, and remixing different paradigms to form a new one. Identify potential advantages and usage scenarios of the new paradigm.\\
& & 3. Study the new paradigm with tool prototypes and example application scenarios. \\


\bottomrule
\end{tabular}}}
\end{table*}

\review{Q6}\revise{\nistart{C4. Reusing, chaining, and innovating atomic paradigms for adapting interaction design to new scenarios}}
Our summarized paradigms and considerations C1-C3 above provide a concise mapping from task requirements to interface affordances.
\review{Q10}\revise{Designers can start from intent clarity and artifact availability to select appropriate paradigms for interface implementation.
For example, they can use the pre-invocation, artifact-grounded paradigm \review{Q6}\revise{(\autoref{tab:cheatsheet}-P5)} for tasks that start from existing artifacts and demand precision.
They can apply post-invocation, prompt-driven paradigms \review{Q6}\revise{(\autoref{tab:cheatsheet} P5-P8)} for exploration goals. 
The should also consider the intent type as C3 mentions.
Crucially, the twelve paradigms can be flexibly combined and chained within a single application, enabling systems to shift fluidly between scaffolding, refinement, and repurposing workflows~\cite{Gmeiner2025, Wang2024g}. 
Effective interfaces thus treat paradigms as composable building blocks rather than rigid templates, supporting diverse user needs.
For example, IntentTagger~\cite{Gmeiner2025} introduces small, atomic intent tags enabling micro-prompting and region-level edits, illustrating chaining and covering multiple paradigms (\eg P2, P4, P5, P7, P9).
Reusing and chaining paradigms are not the end.
Comparing the summarized paradigms and the entire IAI model, we can notice there are plenty of potential atomic paradigms that have not been explored.
In the future, interaction designers may start from existing paradigms to design new ones for innovation, indicating the model's generative power.
}

\section{Usage Scenario} 
\label{sec: case}

The generative power of our proposed interaction-augmented instruction model 
lies in guiding both interaction paradigm and interactive tool innovation in human–GenAI communication~\cite {Beaudouin-Lafon2021GenerativeInteraction}. 
Together with our summarized existing paradigms, it can guide the iterative improvement of tools, facilitate the creation of novel interfaces tailored to emerging user scenarios, and enable generating new interaction paradigms beyond existing ones. 
\review{Q9}\revise{To illustrate the model's generative power, we summarized four potential methods for designers and researchers to apply IAI models and summarized twelve paradigms for various needs, 
as shown in \autoref{tab:design-guidelines}. 

To demonstrate the four methods, we present four usage scenarios below to highlight how our model and paradigms can inspire, structure, and accelerate design decisions in real-world contexts. 
These scenarios were deliberately selected to cover the three fundamental ways in which our model can be used, forming a closed application space of design actions: 
(1) selecting and applying existing paradigms (Usage Scenarios 1 and 3),
(2) refining existing paradigms by adjusting their graph structures (Scenario 2), and
(3) generating new paradigms when existing ones are insufficient for a novel task (Scenario 4). 
The choice of application domains (\eg data science, creativity, and everyday tasks for general users) is intentionally diverse to demonstrate that the IAI model is not domain-bound. 
Rather than exhaustiveness, our goal is to illustrate breadth and accessibility, enabling readers from different communities to understand how the model applies across contexts.
}

\begin{figure*}[t]
  \centering
    \includegraphics[width=\linewidth]{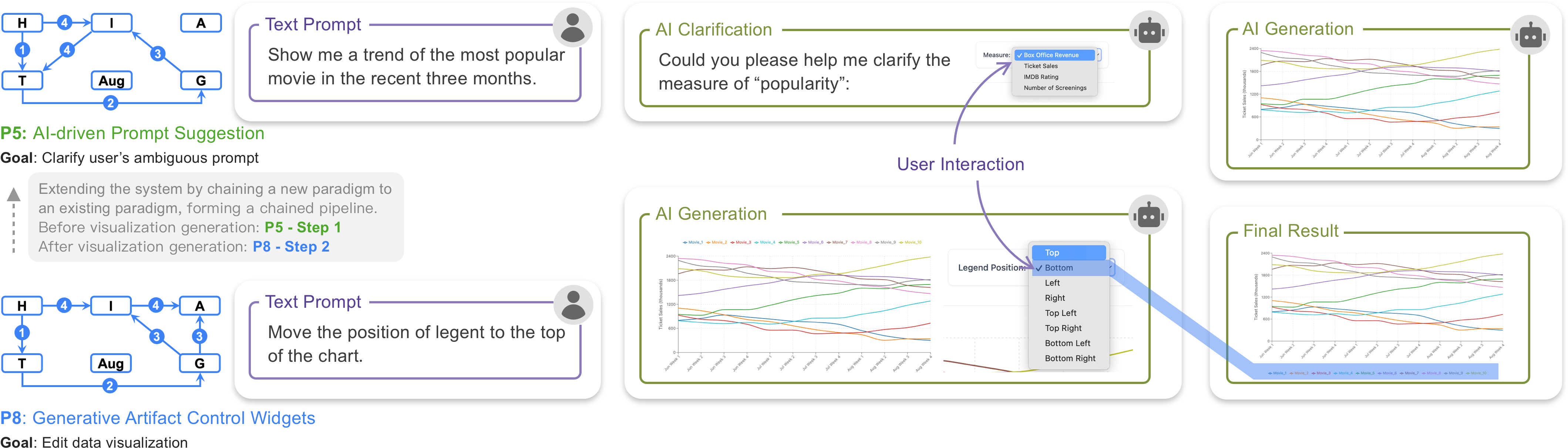}
     \vspace{-10px}
    \caption{
      Usage Scenario 1: Extending Pipelines through Chained Paradigm Graphs.
      For example, DynaVis~\cite{Vaithilingam2024} supports post-generation visualization refinement (P8). By chaining a pre-generation disambiguation paradigm (P5), the system can clarify ambiguous terms before execution, augmenting rather than replacing existing workflows.
    }
\label{fig: usage-scenario-1}
\end{figure*}

\begin{figure*}[t]
  \centering
    \includegraphics[width=\linewidth]{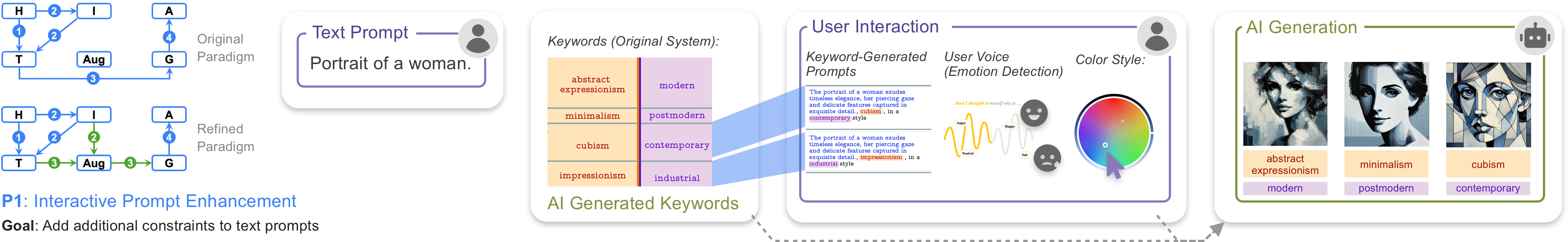}
     \vspace{-10px}
    \caption{
      Usage Scenario 2: Refining Paradigms by Adjusting Graph Structures.
For example, Dreamsheets~\cite{Almeda2024} (P1) can be extended by adding I~$\rightarrow$~Aug, introducing multimodal inputs (\eg voice for emotion, palette for color tone) beyond text prompts. This enables more structured, controllable, and user-steerable exploration.
    }
\label{fig: usage-scenario-2}
\end{figure*}

\subsection{Usage Scenario 1: Extending Pipelines through Chained Paradigm Graphs}
\label{sec: case 1}

One way our model supports innovation is by extending existing human-GenAI interfaces through the chaining of additional atomic paradigm graphs onto their current workflows (\autoref{sec:cross_insights}-C4),  with additional invocations of GenAI. 
This perspective treats current system interfaces not as static endpoints but as expandable foundations, where other paradigms can be strategically layered to address ambiguity and improve alignment

For example, when designing an ideal tool for data analysts to generate and edit visualizations, natural language can ease the burden of translating design requirements into initial visualizations, while graphical interactions can make subsequent adjustments and fine-grained editing more efficient. 
DynaVis~\cite{Vaithilingam2024} (\autoref{fig: paradigm1}-P8) illustrates this synergy by augmenting natural language interfaces with dynamic widgets that support iterative refinement. 
Yet, at the same time, early-stage ambiguity in user intent often remains a challenge.
\autoref{fig: usage-scenario-1} illustrates this, imagine a data analyst at a movie company exploring market trends for the second quarter across U.S. cinemas. 
The analyst might ask: \textit{``Show me a trend of the most popular movie in the most recent three months.''} 
Directly generating data insights to such fuzzy user questions (\eg ``\textit{the most popular}'') may not fully align with the user's intent, frequently leading to time-consuming post-generation refinements.
Using the IAI model and the atomic paradigm graphs, system designers can find and introduce a pre-generation disambiguation paradigm (P5, AI-driven Prompt Suggestion) before P8. 
Instead of immediately generating a visualization, the system could first initiate a clarification step, asking the analyst what ``popular'' should mean in this context, by \textit{ticket sales}, \textit{IMDB rating}, \textit{box office revenue}, or \textit{number of screenings}. 
This step structures the workflow to clarify ambiguous prompts upfront, improving alignment, reducing back-and-forth iterations, and making subsequent widget-driven refinement more precise.

By enabling such chaining, the Interaction-Augmented Instruction model does not replace existing workflows but augments them, making human-GenAI collaboration more user-friendly, accurate, and adaptable to nuanced tasks.


\begin{figure*}[t]
  \centering
    \includegraphics[width=\linewidth]{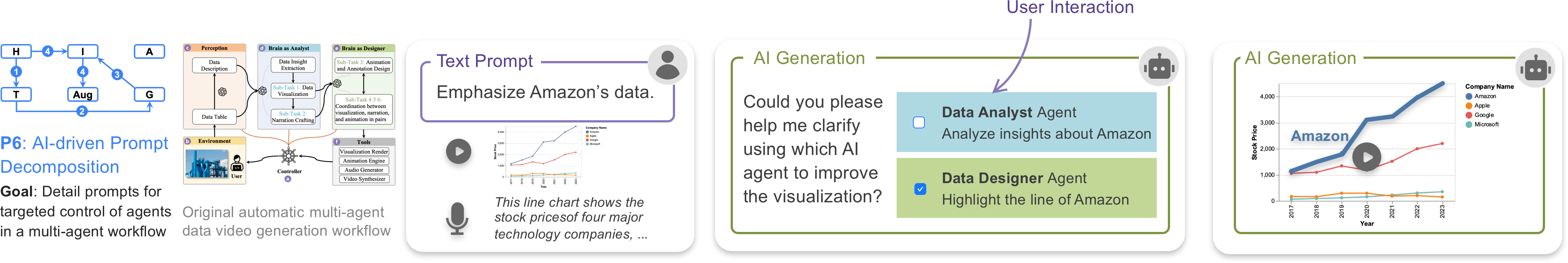}
     \vspace{-10px}
    \caption{
      Usage Scenario 3: Applying the IAI Model to Emerging Scenarios.
For example, in a multi-agent system for animated data videos~\cite{datadirector}, the IAI model guides the integration of user feedback for iterative refinement. This yields a paradigm aligned with P6 and demonstrates the IAI model can serve as a reasoning tool for emerging scenarios.
    }
\label{fig: usage-scenario-3}
\end{figure*}

\subsection{Usage Scenario 2: Refining Paradigms by Adjusting Graph Structures}
\label{sec: case 2}

In addition to strengthening workflows at a macro level by chaining different paradigms, the flexibility of the relations among entities in the IAI model also enables micro-level edits within a single paradigm, without additional invocations of GenAI. 
Such refinements can be realized by adjusting the relations among entities in existing paradigms to improve precision and control.

Take the example of AI artwork creation. 
In media-art contexts, user intent often extends beyond describing the content of an artwork to include more nuanced dimensions, such as conveying emotion, which is an aspect that is difficult to express through text alone and is often better captured through other modalities like voice~\cite{moon2020empathy} and facial expression~\cite{hong2022algorithms}. 
Dreamsheets~\cite{Almeda2024} (\autoref{fig: paradigm0}-P1) provides a strong foundation in this scenario: its spreadsheet-like interactions make prompt refinement and enhancement efficient for rapid iteration in artwork generation and exploration (P1, Interactive Prompt Expansion).
Yet, relying solely on natural language refinement can still be limited in aligning with a user’s intent, as analyzed based on the IAI model in \autoref{sec: pre-call prompt}. 
For example, even a seemingly simple adjustment, such as changing the color style, becomes cumbersome when users must come up with the precise name of the desired style for prompting AI. 
Also, users must rely on texts to describe their nuanced feeling (\eg emotion) to AI.


Using the IAI model, interface designers can systematically identify potential links to enhance Dreamsheets in its corresponding paradigm graph P1. 
Specifically, an addition of a relation from interaction to augmented instruction (I$\rightarrow$Aug) can provide additional intent expression methods beyond natural language.
As shown in \autoref{fig: usage-scenario-2}, in practice, this adjustment could introduce additional interactions for specifying global artwork parameters. 
For instance, after rapidly generating prompt candidates for a ``\textit{portrait of a woman}'' through spreadsheet-like interactions, the user may specify their feeling in creation via voice, where their emotion can be detected.
The user can also adjust the overall color tone using a palette. 
This interface refinement transforms free-form prompt expansion into a more systematic, user-steerable process with an augmented instruction.
It enables more targeted and expressive exploration through appropriate materialization of user intent (\autoref{sec:cross_insights}-C3).


\begin{figure*}[t]
  \centering
    \includegraphics[width=\linewidth]{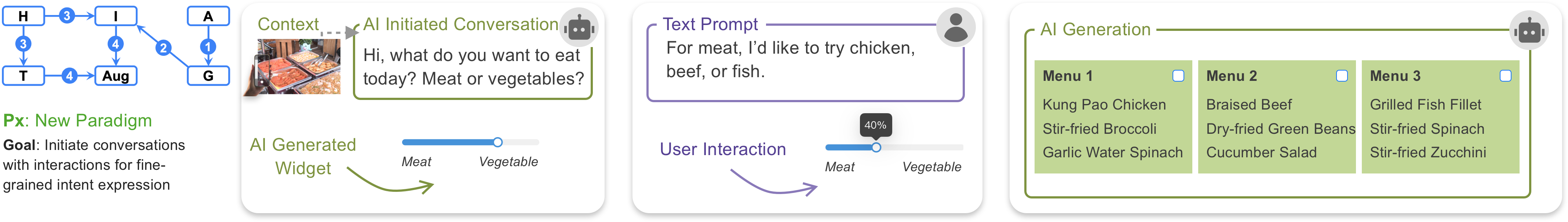}
     \vspace{-10px}
    \caption{
      Usage Scenario 4: Deriving New Atomic Paradigm Graphs.
      For example, modifying P11 suggests a new paradigm where AI proactively initiates interaction. In a canteen scenario, an assistant proposes menu options through contextual sensing and interactive widgets, illustrating how paradigm modifications can inspire novel applications.
}
\label{fig: usage-scenario-4}
\end{figure*}

\subsection{Usage Scenario 3: Applying the IAI model to Emerging Scenarios}
\label{sec: case 3}

A distinctive strength of the IAI model lies in its generative capacity: by formalizing six core entities and their relations, the model can guide system designers in deriving atomic paradigm graphs for new human-GenAI applications. 
This process is especially valuable in the emerging scenario of multi-agent workflows, which often fall short of supporting fine-grained iterative refinement. 

Consider Data Director~\cite{datadirector}, a multi-agent system that automatically generates animated data videos from data tables with agent roles such as Data Analyst and Designer (\autoref{fig: usage-scenario-3}). 
Although such end-to-end automation can rapidly deliver initial outputs, its results often require follow-up refinement by humans. 
For example, in the case study of stock price analysis, after generating a data video with an animated line chart of multiple companies and corresponding narration, the user may issue a follow-up instruction: \textit{``Emphasize Amazon's data.''} 
How might the system designer extend the existing multi-agent workflow to accommodate this new requirement?

Using the IAI model, the system designer begins by identifying the relevant entities: Human (H), Text Prompt (T), Generative AI (G), and Interaction (I), where Generative AI is to understand the user's fine-tuning instruction and refine the initial prompt in the multi-agent system. 
Next, the system designer specifies the relations required to fulfill the user's intent. 
First, the human issues the request (H~$\rightarrow$~T). In this multi-agent system, producing a data video involves multiple tasks such as extracting insights, generating visualizations, crafting narration, and creating animations or annotations. Accordingly, a single instruction may map to different agent actions, such as the Designer agent animating Amazon’s line or the Data Analyst agent focusing on Amazon’s insights to craft narration.
To ensure agent-specific accuracy, GenAI decomposes the text prompt and proposes candidate actions by different agents (T~$\rightarrow$~G, G~$\rightarrow$~I), and the human then reviews and confirms these suggestions (H~$\rightarrow$~I). 
Thus, at this scenario, the interaction occurs \textit{after} calling GenAI (\autoref{sec:cross_insights}-C1) and involves a text-only starting resource (\autoref{sec:cross_insights}-C2), consistent with the ``Post-invocation, Prompt-only'' category (\autoref{sec: post-call prompt}). 
Finally, the confirmed interaction forms a new augmented instruction (I~$\rightarrow$~Aug), which is passed back for another round of GenAI calling (chained by a follow-up paradigm). 

This design process yields a paradigm that aligns with P6, AI-driven Prompt Decomposition. 
By explicitly modeling entities and relations, the IAI framework helps system designers localize where and how human input should occur, and how to derive paradigms to support iterative communication between humans and AI. 
More broadly, this case demonstrates how the IAI model can serve as a reasoning tool for emerging scenarios: starting with entities, constraining relations along the two axes of interaction timing and user resources (\autoref{tab: paradigm}), and applying cross-paradigm insights (\autoref{sec:cross_insights}) to guide concrete design choices. 

\review{Q10}\revise{
Notably, while this scenario illustrates how the IAI model can support multi-agent workflow design, multi-agent systems are only one application context. The model does not formalize communication among multiple AI agents or include multiple \textbf{G} nodes within a single atomic paradigm. Each paradigm intentionally centers a single \textbf{G} node, and multi-agent workflows are expressed by chaining atomic paradigms if \textbf{G} is called multiple times for different purposes. Extending the model to represent shared or distributed \textbf{Aug} across agents is a promising future direction.
}

\subsection{Usage Scenario 4: Deriving New Atomic Paradigm Graphs}
\label{sec: case 4}

Beyond applying the IAI model to new usage scenarios, its generative capacity also enables HCI researchers to hypothesize and explore new paradigms of human-GenAI collaboration. 
While the twelve atomic paradigm graphs we distilled capture a representative set of existing practices, they do not exhaust the design space. 
In this case, we take the reverse perspective: 
rather than deriving paradigms from scenarios, we start by modifying existing paradigm graphs to see how such changes may give rise to new paradigms and, in turn, novel application scenarios. 

Consider the paradigm P11, Artifact-driven Prompt Enhancement (\autoref{fig: paradigm2}-P11), where GenAI initiates interactions based on information from artifacts (contextual information) and human prompts. 
Contextual information is critical in many scenarios, such as embodied AI in everyday life. 
But instead of a human initiating the conversation based on context like P11 (starting from H~$\rightarrow$~T), what if the \textbf{AI initiates the conversation proactively}, as illustrated in \autoref{fig: usage-scenario-4} (left) and exemplified in \autoref{fig: paradigm2}-P9 and P10 (starting from A~$\rightarrow$~G)?

Assume an individual enters a canteen and inspects the available dishes, the human–AI conversation is not initiated by the human through an explicit prompt (\autoref{fig: usage-scenario-4}). 
Instead, an embodied AI assistant (\eg embedded in AR glasses or a mobile app~\cite{jiang2025dietglance, Wang2024i}) proactively perceives the environment (\autoref{sec:cross_insights}-C2), cross-references it with the individual’s dietary history, and initiates the interaction: \emph{``\textit{Hi, what do you want to eat today? Meat or vegetables?}''}
Alongside this query, the system also generates an interactive widget (\eg a slider) for specifying a preferred proportion of vegetables versus meat. 
Suppose the user selects 60\% vegetables and 40\% meat, and adds: ``\textit{For meat, I’d like to try chicken, beef, or fish.}''
The AI then integrates this input with the detected canteen offerings to recommend a personalized list of top dishes. 
%
This new graph and usage scenario foregrounds \textbf{AI-initiated, context-aware interaction}, expanding the design space toward more proactive and situated human-GenAI collaborations (\autoref{sec:cross_insights}-C4), and more broadly, opening up fundamentally new paradigms of communication.

\section{Discussion}

This section reflects our research  (Sec.~\ref{sec:reflection}) and outlines future directions (Sec.~\ref{sec:future_work}).

\subsection{Reflection}\label{sec:reflection}


\noindent\textbf{Why do we need instruction-augmented interaction?}
We model the interplay between prompts and interactions because GenAI systems increasingly rely on both, yet lack a framework to make their complementarity explicit.
The IAI model and the twelve atomic paradigms articulate a core claim: prompts and focused interactions are complementary (not interchangeable) modes for externalizing user intent. Treating Augmented Instruction (Aug) as the explicit input to generative models foregrounds a functional separation: Text Prompt (T) supplies high-level, abstract goals; Interaction (I) supplies precise, referential constraints and grounding. 
There are three closely related reasons this interplay is necessary. First, generative models map underspecified instructions to a broad set of plausible outputs~\cite{Riche2025}; interaction signals collapse referential ambiguity and materially improve the likelihood of a targeted, single-turn outcome~\cite{rost2025reclaiming}. Second, interactions encode provenance and manipulable constraints that support fine-grained control properties that language alone cannot reliably provide at scale~\cite{Shneiderman1997}.
Finally, the IAI model brings human–GenAI communication closer to human–human interaction. In practice, people rarely rely on language alone; they complement speech with gestures, sketches, and shared artifacts to ground intent. By mirroring these multimodal practices, IAI reduces ambiguity and enriches expression, pointing toward GenAI systems that act as more natural collaborators in human–AI co-creation.

\vspace{\baselineskip}
\noindent\textbf{How should we apply interaction-augmented instructions?}
Adopting interaction-augmented instructions implies several foundational shifts in human–AI practices. 
First, expertise will shift from prompt engineering to instruction design. Users should compose reusable and multimodal instructions rather than optimizing text alone, while tools need to expose these capabilities with clear discoverability to support effective use~\cite{Dontcheva}.
Second, cognitive load and UX trade-offs must be managed: interactions can reduce text specification but add interface complexity, so designers should prioritize low-friction affordances (\eg defaults, progressive disclosure, previews) to minimize overhead~\cite{Subramonyam2023}. 
Third, tool evaluation can broaden beyond artifact quality to include axes that follow directly from the IAI distinction, such as correctness of local changes tied to interactions (referential fidelity), rounds to satisfactory output (convergence cycles), ease of steering behavior (controllability), and  user comprehension of influences (provenance clarity). 
\vspace{\baselineskip}
\noindent\textbf{What can our model contribute to interaction-augmented instructions?}
The IAI model contributes to both engineering practice and HCI theory-building.
\review{Q2}\revise{It provides greater descriptive, discriminative, and generative power than methods in prior work, such as Gao~\etal~\cite{Gao2024} and Shen~\etal~\cite{Instructions}, by introducing two innovations.
First, the model explicitly encodes \textit{information flow} in the paradigm graph. Each edge denotes a precise link between entities, making differences in timing, source, and composition of instructions formally visible.
Another key contribution is treating \textbf{Aug} as an explicit instruction to GenAI for the expressiveness of the model. Without it, designs that combine text prompts, interactions, and artifacts in different ways collapse into the broad ``non-text interactions'' category. 
It cannot systematically distinguish important differences, such as that between P1 and P2.
With \textbf{Aug}, these differences become structurally visible, varying how entities feed into GenAI, and in what order, systematically yielding distinct paradigms. 
This mechanism directly enables both our comparative taxonomy and its generative extension, where new interfaces emerge from principled recombinations of information flow patterns.}

For practitioners, paradigm graphs serve as a design language that makes implicit design trade-offs explicit. Developers can compare alternatives, identify unexplored regions of the design space, and refine existing interfaces by recombining atomic paradigms. This resonates with prior calls to move beyond ad-hoc demonstrations toward systematic design frameworks in HCI~\cite{Beaudouin-Lafon2004}.
For researchers, our model complements prior taxonomies~\cite{Gao2024, Cook2025} and principle-based approaches~\cite{Riche2025} by introducing a formal representational structure that captures both entities and relations in human–GenAI workflows. 
\re{
For example, the three principles proposed in AI-Instruments~\cite{Riche2025} can be reinterpreted in IAI as properties of interaction flows rather than standalone design goals.
\textit{Reification} corresponds to whether and how intent is externalized into explicit intermediate representations (most notably \textbf{Aug}) that persist beyond a single interaction and can be inspected, reused, or chained.
\textit{Reflection} maps to interaction flows in which multiple alternatives are explicitly surfaced and negotiated over time, often through post-invocation paradigms where model outputs serve as anchors for iterative clarification.
\textit{Grounding} is realized through paradigms that bind instructions to concrete artifact scopes (\eg selected artifacts, regions, or prior outputs), thereby constraining interpretation through reference rather than description.
Rather than treating these principles as orthogonal design heuristics, IAI shows how they manifest repeatedly across different atomic paradigms depending on interaction timing, available resources, and control flow.
This shift reframes human–GenAI interaction from early-stage guideline analysis toward the concrete design and enactment of interaction structures, thereby enabling both actionable system designs and principled exploration of new paradigms.
}
Such unified formalization enables cumulative comparison across systems and provides a substrate for analytical and generative methods, similar to how prior models advanced earlier eras of HCI~\cite{card1986model, Beaudouin-Lafon2000, Card1999}.

\subsection{Future Work}\label{sec:future_work}

\nistart{Model: Extending the IAI Model}
The IAI model intentionally abstracts at a high level to capture the interplay between text prompts and interactions. 
This abstraction aims for generality, but extensions can enrich the model for more fine-grained analysis of specific design questions.
At the entity level, entities can be refined or expanded to capture richer system dynamics. For instance, as noted in \autoref{sec:model_entity},  while \textit{context} is currently subsumed under the \textit{artifact} entity, making it explicit would support the analysis of how retrieval-augmented systems use user provenance. 
In addition, the ``human'' entity could generalize into an \textit{actor} role, instantiated by either humans or AI agents, enabling representation of mixed-initiative or multi-AI agent workflows~\cite{Subramonyam2023, Li2023c}.
At the level of relations, finer distinctions can also sharpen analysis. For example, the link from \textit{interaction} to \textit{augmented instruction} manifests differently across tools, ranging from sliders and widgets to sketching or structured graph editing. Making such variations explicit would not only capture existing diversity but also open design opportunities, such as composable augmented instruction libraries for domain-specific workflows~\cite{Setlur2020}, adaptive timing of system responses~\cite{Gajos2006}, or model-assisted widget generation~\cite{Chen2025b}.

\vspace{\baselineskip}
\nistart{Paradigm: Iterative Expansion of Paradigms and Next-Generation Scenarios}
The twelve atomic paradigms are not exhaustive but serve as a core library for iterative growth. Expansion occurs iteratively along multiple paths, as illustrated by our four usage scenarios (\autoref{sec: case}). 
Paradigms can be extended by \textit{chaining} additional graphs onto existing workflows (Scenario 1), \textit{refined} through structural adjustments to better materialize user intent (Scenario 2), \textit{applied} to emerging domains to guide design reasoning (Scenario 3), or even \textit{modified} to hypothesize entirely new paradigms that inspire novel applications (Scenario 4).
These scenarios illustrate how atomic paradigms function as both descriptive lenses and generative building blocks, capturing current practices while revealing underexplored design subspaces.
Looking ahead, next-generation contexts will further expand the paradigm space. XR brings gaze and embodied gestures~\cite{Hong}; cross-device and multimodal workflows demand seamless orchestration~\cite{Lee2021c}; adaptive systems restructure interfaces dynamically~\cite{Xu2020}; and affective computing raises questions of how emotion should shape interaction. 
Our model is extensible to these futures by augmenting paradigm graphs with new entities and relations, offering a systematic scaffold for both incremental refinement and paradigm-level innovation.

\vspace{\baselineskip}
\nistart{Interface: From Paradigms to Generative Design}
The IAI model is both descriptive and discriminative, as well as generative. It decomposes hybrid system interfaces into atomic paradigm graphs, while discriminatively highlights structural differences across interfaces.
Generatively, paradigm graphs serve as design blueprints~\cite{Beaudouin-Lafon2021GenerativeInteraction}, guiding the refinement of existing interfaces and the creation of new ones.
Our usage scenarios (\autoref{sec: case}) illustrate this generative role in practice, showing how paradigm graphs can inspire novel interface designs for various application scenarios.
They demonstrate how paradigm graphs not only capture existing practices but also scaffold systematic innovation.
Beyond manual application of these paradigm graphs, we envision that it can also inspire UI generation.
For example, based on user intent clarity and the availability of artifacts (\autoref{sec:cross_insights}), AI models can select suitable paradigms and generate adaptive interfaces to guide users for next round of communication.

\vspace{\baselineskip}
\revise{\nistart{Limitations}
Our contributions should be interpreted within several boundaries. 
First, the IAI model and the analyzed atomic paradigms currently focus on single-user, single-agent interactions and do not yet formalize coordination among multiple GenAI nodes or the construction of shared augmented instructions.
Second, the taxonomy is derived from a UI-centric corpus of 66 tools, which, although diverse, cannot fully represent the rapidly evolving GenAI ecosystem. 
Third, validation relies on qualitative synthesis and illustrative case studies rather than empirical user or designer studies. Future work could include controlled evaluations and real-world deployments to assess practical adoption and impact.

}

\section{Conclusion}
We introduced the Interaction-Augmented Instruction (IAI) model that formalize how natural language prompts and GUI interactions jointly shape human–GenAI communication. 
With the IAI model, we summarized twelve atomic paradigms based on existing tools, which provide reusable abstractions that enable systematic characterization, comparison, and innovation in interface design. 
Our usage scenarios demonstrate how this model and the extracted twelve atomic paradigms bridge conceptual models with actionable design choices, supporting refinement of existing tools and exploration of new interaction spaces. 
The paradigms and usage scenarios jointly verify that our proposed IAI model have sufficient descriptive, discriminative, and generative power to model the interplay of prompts and GUI interactions.
In future, we plan to further explore this research direction through extending the model to finer granularity, expanding the collection of paradigms, and exploring more usage scenarios for the model and paradigms, such as generative UI.

\begin{acks}
This research is part of the AFMR collaboration supported by Microsoft Research.
This work is partially supported by Hong Kong RGC GRF grant (16218724).
\end{acks}

\bibliographystyle{ACM-Reference-Format}

\bibliography{main}

\clearpage

\end{document}